\documentclass[letterpaper,12pt]{JHEP3}
\usepackage{amsfonts}
\usepackage{amsmath,amssymb}
\usepackage{latexsym,stmaryrd}
\usepackage{dsfont}
\usepackage{amsthm}
\usepackage{pstricks}
\usepackage{pst-plot}
\usepackage{epsfig}
\usepackage{subfigure}
\usepackage{cite}
\usepackage[english]{babel}

\newcommand{\eq}{\begin{equation}}
\newcommand{\feq}{\end{equation}}
\newcommand{\eqn}{\begin{eqnarray}}
\newcommand{\feqn}{\end{eqnarray}}
\newcommand{\arr}{\begin{eqnarray*}}
\newcommand{\farr}{\end{eqnarray*}}

\newcommand{\lp}{\left(}
\newcommand{\rp}{\right)}
\font\mybb=msbm10 at 12pt
\def\bb#1{\hbox{\mybb#1}}

\def\bR {\bb{R}}

\def\ga{\gamma}

\newcommand{\e}{\operatorname{e}}
\newcommand{\de}{\operatorname{d}\!}

\newlength{\diameter}
\newcommand{\olett}[1]{
  \ensuremath{
  \text{$\settowidth{\diameter}{$\bigcirc$}%
  \bigcirc%
  \hspace{-0.5\diameter}%
  \makebox[0pt][c]{$\scriptstyle{#1}$}%
  \hspace{0.5\diameter}$}}}    

\newlength{\unit}
\psset{xunit=\unit,yunit=\unit,runit=\unit}
\newlength{\linew}
\psset{linewidth=\linew}

\newgray{mygray}{0.6}

\title{Black strings in AdS$_5$}

\author{Alice Bernamonti,$^a$ Marco M.~Caldarelli,$^b$ Dietmar
Klemm,$^{ac}$ Rodrigo Olea,$^c$ Christoph Sieg$^c$ and Emanuele Zorzan$^{ac}$ \\
$^a$ Dipartimento di Fisica dell'Universit\`a di Milano, \\
\hspace*{0.15cm} Via Celoria 16, I-20133 Milano. \\
$^b$ Departament de F\'{i}sica Fonamental, \\
\hspace*{0.15cm} Universitat de Barcelona, \\
\hspace*{0.15cm} Diagonal, 647, 08028 Barcelona, Spain. \\
$^c$ INFN, Sezione di Milano, Via Celoria 16, I-20133 Milano. \\
E-mail: \email{alice.bernamonti@libero.it, caldarelli@ub.edu,
dietmar.klemm@mi.infn.it, rodrigo.olea@mi.infn.it, csieg@mi.infn.it,
emanuele.zorzan@mi.infn.it}}
\preprint{IFUM-898-FT}

\abstract{We present non-extremal magnetic black string solutions in five-dimensional
gauged supergravity. The conformal infinity is the product of time and S$^1$
$\times$ ${\cal S}_h$, where ${\cal S}_h$ denotes a compact Riemann surface of
genus $h$. The construction is based on both analytical and numerical techniques.
We compute the holographic stress tensor, the Euclidean action and the conserved
charges of the solutions and show that the latter satisfy a Smarr-type formula.
The phase structure
is determined in the canonical ensemble, and it is shown that there is a first
order phase transition from small to large black strings, which disappears above
a certain critical magnetic charge that is obtained numerically.
For another particular value of the magnetic charge, that
corresponds to a twisting of the dual super Yang-Mills theory,
the conformal anomalies coming from the background curvature and those arising
from the coupling to external gauge fields exactly cancel. We also obtain
supersymmetric solutions describing waves propagating on extremal BPS magnetic
black strings, and show that they possess a Siklos-Virasoro reparametrization
invariance.}


\begin{document}

\section{Introduction}
\label{intro}

Four dimensional black holes are well understood. The uniqueness theorems ensure that for a given set of asymptotic charges, a unique black hole phase exists and belongs to the Kerr-Newman family of solutions. Things change drastically in higher dimensions. The discovery of a five dimensional black ring solution \cite{Emparan:2001wn} has shown that uniqueness is violated. This leads to interesting phase diagrams, where a phase transition between Myers-Perry black holes \cite{Myers:1986un} and black rings occurs as one increases the angular momentum of the system.
The phase diagram becomes much richer if one spacelike dimension is compactified. The uniform black string solution, that can be constructed as a direct product of the Schwarzschild solution times the compact dimension, suffers from a long wavelength gravitational instability, the Gregory-Laflamme instability \cite{Gregory:1993vy}, but becomes stable above a critical mass. At this critical mass new static non-uniform strings emerge, with the same horizon topology of the black string, but without the translational symmetry along the circle. Another phase is given by Kaluza-Klein black holes, black holes with $\text{S}^3$ horizon topology localized on the circle. This localized black hole phase meets the non-uniform string phase in a topology transition point, the merger point. The dynamics of the decay of these strings is still unclear and opens interesting questions related to the cosmic censorship. Other phases coexist in the full diagram, and phases where Kaluza-Klein bubbles are attached to black holes and multi-black hole configurations have been studied in the literature, showing a continuous non-uniqueness of static classical black hole solutions for given asymptotic charges (see \cite{Harmark:2007md,Kol:2004ww} for reviews on the subject).

The study of black holes in presence of a negative cosmological constant is of particular interest in the context of the AdS/CFT correspondence \cite{Maldacena:1997re} since their thermodynamics opens the opportunity to shed some light on the non-perturbative aspects of certain field theories and on their thermodynamical phases. For example, the Hawking-Page transition \cite{Hawking:1982dh} between thermal AdS$_5$ and the Schwarzschild-AdS$_5$ black hole corresponds to a thermal phase transition from a confining to a deconfining phase of $D=4$, ${\mathcal N}=4$ SYM on ${\mathbb R}\times\text{S}^3$ \cite{Witten:1998zw}, while the phase structure of Reissner-Nordstr\"om-AdS black holes, similar to that of the van der Waals-Maxwell liquid-gas system, is connected to the study of the SYM theory coupled to some background R-symmetry current \cite{Chamblin:1999tk,Chamblin:1999hg}.
Moreover, if one compactifies the SYM theory on a Scherk-Schwarz $\text{S}^1$, the resulting low energy dynamics is that of a $2+1$ dimensional Yang-Mills theory that undergoes a deconfining phase transition at finite temperature. Lumps of deconfined plasma have then an effective description in terms of fluid dynamics. The phases of this fluid, studied in \cite{Aharony:2005bm,Lahiri:2007ae}, are directly related to the black hole phases of the gravitational description\footnote{See \cite{Bhattacharyya:2007vs} for recent work on the subject.}. The existence of plasmarings suggests the existence of asymptotically AdS$_5$ black rings, yet to be discovered. Further independent evidence has been obtained by the authors of \cite{Kunduri:2006uh}, that found supersymmetric black rings in AdS$_5$ presenting a conical singularity, which plausibly will disappear out of extremality by balancing the forces.

If one allows for more general topology of the boundary, locally asymptotically AdS black holes with different horizon topology are permitted \cite{Mann:1997iz,Vanzo:1997gw}. The horizon of these so-called topological black holes can be any Einstein space with positive, negative or vanishing curvature \cite{Birmingham:1998nr}. In this article we will focus on minimal gauged supergravity in five dimensions, on locally asymptotically AdS$_5$ spacetimes with one non-trivial $\text{S}^1$ cycle at infinity. The boundary where the CFT lives has then the topology ${\mathbb R}_t\times{\mathcal S}\times\text{S}^1$ where $\mathcal S$ can be a two-sphere $\text{S}^2$, the Lobatchevski plane ${\mathbb H}^2$ or the Euclidean space E$^2$. Not much is known on the black hole phases in this case; uniform, neutral black strings with $\text{S}^2\times\text{S}^1$ topology have been found by Copsey and Horowitz in \cite{Copsey:2006br} and then generalized to higher dimensions and arbitrary $\mathcal S$ by Mann, Radu and Stelea in \cite{Mann:2006yi}. Electric charge and angular momentum were included in \cite{Brihaye:2007vm}, while non-abelian black string solution (but also abelian $U(1)$) were obtained in \cite{Brihaye:2007ju}. These solutions are typically numerical, since in presence of a cosmological constant the black strings cannot be constructed by taking a direct product with a circle, and the differential equations they verify cannot be solved analytically. There are exact supersymmetric magnetically charged solutions \cite{Chamseddine:1999xk,Klemm:2000nj} with horizon topology $\text{S}^2\times\text{S}^1$ or ${\mathbb H}^2\times\text{S}^1$, but their magnetic charge is quantized in terms of the AdS radius and there is no smooth limit connecting them to the uncharged strings.

In this article we will obtain the general family of magnetically charged black strings in AdS$_5$, connecting the supersymmetric to the uncharged ones. We construct them by matching numerically the near-horizon expansion of the metric to their asymptotic Fefferman-Graham expansion \cite{FG}. Furthermore, for some special cases, we are able to find their exact analytical expressions. We compute their masses and tensions, using the counterterm prescription \cite{Balasubramanian:1999re}, and fix the vacuum energy with the Kounterterm procedure \cite{Olea:2005gb,Olea:2006vd}. If the magnetic charge assumes a certain value given in terms of the inverse gauge 
coupling constant (corresponding to a twisting of the dual super Yang-Mills theory), 
the conformal anomalies coming from the background curvature and those arising from 
the coupling to external gauge fields exactly cancel\footnote{This was already noticed in \cite{Brihaye:2007ju}.}.
The study of their thermodynamics shows in the canonical ensemble (fixed magnetic charge) a van der Waals-Maxwell phase structure similar to the one of the Reissner-Nordstr\"om-AdS black holes \cite{Chamblin:1999tk,Chamblin:1999hg} and electrically charged black strings in AdS$_5$ \cite{Brihaye:2007vm}: for small magnetic charges, we find two coexisting black string phases separated by a first order phase transition, which disappears at a critical point as we increase the charge.

Performing a double analytic continuation, these solutions describe static, magnetically charged bubbles of nothing in AdS$_5$. For vanishing cosmological constant, the Kaluza-Klein bubbles can have arbitrarily negative energy, implying an instability of this vacuum sector of Einstein's equations. For asymptotically AdS solutions, the existence of a stable ground state in the dual field theory description implies a lower bound on the mass of these solutions. This has led to the formulation of the positive energy conjecture for locally asymptotically AdS spacetimes \cite{Horowitz:1998ha}, that has been discussed in \cite{Copsey:2006br} in presence of one compact direction on the boundary. We show that the static magnetically charged bubbles exist for any size of the S$^1$ at infinity in contrast to the uncharged case where bubbles exist only below a critical size of the S$^1$. Moreover, the quantum phase transition occuring in the strongly coupled gauge theory as one varies the size of the S$^1$ \cite{Copsey:2006br} becomes a quantum phase transition between the vacua dual to the small and large bubbles of nothing, and disappears for magnetic charges above a critical charge. We expect the lowest energy bubble to have the lowest energy among all solutions sharing the same asymptotic structure, as was conjectured for the uncharged one. It would be interesting to check this by some explicit calculation.

Finally, we turn the attention to supersymmetric black strings and generalize the exact supersymmetric black string solution of \cite{Chamseddine:1999xk,Klemm:2000nj} in two ways. First, we include electric charge, and find new exact solutions representing supersymmetric dyonic black strings with ${\mathbb H}^2\times\text{S}^1$ topology of the horizon. Then, we construct BPS solutions corresponding to magnetic black strings with waves propagating along them. We show that the latter, enjoy a large Siklos-Virasoro reparametrization invariance.

The outline of this article is as follows. In section \ref{sec:magnetic} we  derive the new non-extremal magnetic black string solutions in five-dimensional minimal gauged supergravity, and present along with the numerical results some new exact solutions.
Section \ref{properties} is devoted to the properties of the solutions. We compute the conserved quantities, the conformal anomaly of the dual CFT, and we study their thermodynamics, showing the emergence of a van der Waals-Maxwell phase structure. In section \ref{waves} we find new exact supersymmetric solutions with $\mathbb{H}^2\times\text{S}^1$ horizon topology, generalizing the known BPS magnetic string to non-vanishing electric charge and momentum waves in addition to the quantized magnetic charge. We also show that it enjoys a large Siklos-Virasoro reparameterization invariance. We conclude in section \ref{finalrem} with some final remarks. The first terms of the Fefferman-Graham expansion for the magnetic black strings are given in appendix \ref{appendix}.

\section{Non-extremal magnetic black string solutions}
\label{sec:magnetic}
\subsection{Action principle and field equations}

The theory we shall be considering is minimal gauged supergravity in five 
dimensions,
with bosonic action
\begin{equation} \label{action}
I_0 = \frac1{4\pi G} \int_{\cal M}\left[\left(\frac R4 + 3g^2\right)\star 1 
- \frac 12
F\wedge\star F - \frac 2{3\sqrt 3}F\wedge F\wedge A\right] - \frac 1{8\pi G}
\int_{\partial{\cal M}}\textrm{d}^4 x \sqrt{-\gamma} K\,,
\end{equation}
where $R$ is the scalar curvature and $F=\de A$ is the field strength of the 
U$(1)$ 
gauge field. $K$, appearing in the Gibbons-Hawking term, represents the
trace of the extrinsic curvature of the boundary defined as
\begin{equation}
K_{ij} = -\frac 12(\nabla_i n_j + \nabla_j n_i)\,,
\end{equation}
with $n^j$ denoting the outward pointing normal vector to the boundary and
$\gamma_{ij}$ the induced metric. The equations of motion following from
\eqref{action} are
\begin{equation}
R_{\mu\nu} = 2F_{\mu\rho}{F_{\nu}}^{\rho} - \frac13 g_{\mu\nu}(F^2 + 12 g^2)\,,
             \label{Einstein}
\end{equation}
\begin{equation} \label{maxwell}
\de\star F + \frac 2{\sqrt 3}F\wedge F = 0\,,
\end{equation}
where $F^2 = F_{\mu\nu}F^{\mu\nu}$.

In order to find non-extremal string solutions, we choose the ansatz
\begin{equation} \label{ans-metr}
\de s^2 = -\e^{2V}\de t^2 + \e^{2T}\de z^2 + \e^{2U}\de r^2 + r^2\de\Omega_k^2\,,
\end{equation}
where $V,T,U$ are functions of $r$ only and $\de\Omega_k^2$ denotes
the metric of a two-manifold $\mathcal{S}$ of constant Gaussian
curvature $k$. Without loss of generality we can restrict ourselves to the
cases $k=0,\pm 1$; $\mathcal{S}$ is a quotient space of the universal
coverings $\text{S}^2$ ($k = 1$), ${\mathbb H}^2$ ($k=-1$) or E$^2$ ($k=0$).
Explicitly, we choose
\begin{equation}
\de\Omega_k^2 = \de\theta^2 + S(\theta)^2 \de\varphi^2\,,
\end{equation}
with
$$
 S(\theta)= \left\{
\begin{array}{r@{\;,\quad}l}
\sin\theta & k=1\,,\\
\theta & k=0\,,\\
\sinh\theta & k=-1\,.
\end{array}
\right.
$$
We also assume that the direction $z$ is periodic with period $L$.
In five dimensions, strings can carry magnetic charge. So we take the
magnetic ansatz
\begin{equation}\label{F}
F_{\theta\varphi} = k q S(\theta),\qquad 
A_{\varphi} = k q \int S(\theta)\de\theta
\end{equation}
for the U(1) gauge field. With this choice, the Maxwell equations 
\eqref{maxwell}
are trivially satisfied. Plugging the line element \eqref{ans-metr} into the
Einstein equations \eqref{Einstein}, yields the following set of
coupled ordinary differential equations:
\begin{align}
& \e^{-2U}\left[V'T' - V'U' + V'' + (V')^2 + \frac{2V'}r\right] =
\frac 23\left(\frac{kq}{r^2}\right)^2 + 4g^2\,, \label{eq1} \\
& \e^{-2U}\left[V'T' - T'U' + T'' + (T')^2 + \frac{2T'}r\right] =
\frac 23\left(\frac{kq}{r^2}\right)^2 + 4g^2\,, \label{eq2} \\
& \e^{-2U}\left[V'' + (V')^2 - V'U' + T'' + (T')^2 - T'U' - \frac{2U'}r\right] =
\frac 23\left(\frac{kq}{r^2}\right)^2 + 4g^2\,, \label{eq3} \\
& -\e^{-2U}\left[r(V' + T' - U') + 1\right] + k = \frac 43\left(\frac{kq}r\right)^2
- 4g^2r^2\,. \label{eq4}
\end{align}
We now define $F=V+T$ and $G=V-T$ and consider the difference between 
\eqref{eq1} and \eqref{eq2}, which, after integration, gives
\begin{equation}
G' = \frac{\mu}{r^2}\e^{U-F}\,, \label{eqG'}
\end{equation}
where $\mu$ is an integration constant. Note that for extremal solutions
one has $T=V$, which implies $G=0$ and hence $\mu=0$. Thus $\mu$ can
be interpreted as a non-extremality parameter.
Equation \eqref{eq4} can be rewritten in the form
\begin{equation}\label{eqF'}
F' = \e^{2U}f + U' - \frac 1r\,,
\end{equation}
with
\begin{equation}
f(r) = 4g^2r + \frac kr - \frac{4(kq)^2}{3r^3}\,.
\end{equation}
Using \eqref{eqF'} in the sum of \eqref{eq1} and \eqref{eq2} yields an equation
for $U$ only,
\begin{equation} \label{eqU}
\e^{2U} f^2 + 3 U' f + f' + U'' \e^{-2U} + \frac{U'}r \e^{-2U} = 8g^2 +
\frac 43\left(\frac{kq}{r^2}\right)^2\,,
\end{equation}
which can be written in the more convenient form
\begin{equation}\label{eqy}
y''y + 3 y'f + \frac 1r y'y - (y')^2 - 2y\left[f' - 8g^2 -
\frac 43\left(\frac{kq}{r^2}\right)^2\right] = 2 f^2
\end{equation}
by defining $y=\e^{-2U}$. The only equation that we did not use up to now is
\eqref{eq3}. Subtracting the sum of \eqref{eq1} and \eqref{eq2} from 
\eqref{eq3}, and using \eqref{eqG'} and \eqref{eqF'}, one obtains
\begin{equation} \label{eqF}
\frac{\mu^2}{2r^4}\e^{-2F} = \e^{-2U}\left[\frac{3U'}r + \frac 12(U')^2 
- \frac 3{2r^2}
\right] + \frac 12 \e^{2U}f^2 + fU' + \frac k{r^2} 
- 2\left(\frac{kq}{r^2}\right)^2\,.
\end{equation}
Notice that in the extremal case $T=V$, $\mu=0$, this implies a first order
differential equation for 
$U$.\footnote{This first order equation implies \eqref{eqU}.}
It is straightforward to show that solving \eqref{eqF} for F, deriving
with respect to $r$, and using \eqref{eqU}, leads to \eqref{eqF'}. Thus, in
order to find the complete solution, one first solves \eqref{eqy} to
get $y$. Plugging this into \eqref{eqF} gives then $F$, and finally
\eqref{eqG'} yields $G$. Then all the field equations are
satisfied. We have thus decoupled completely the field equations and
reduced the problem to solving a non-linear ordinary second order
differential equation. Unfortunately, solving \eqref{eqy} is a quite
formidable task, and we did not succeed in finding the most general solution,
so that in the general case we have to resort to numerical techniques.
Nevertheless, it is possible to obtain some particular exact
solutions, which we will discuss in section \ref{exact-sol}.

\subsection{Asymptotics}\label{as}

At large $r$, the functions appearing in the metric admit the
Fefferman-Graham expansions \cite{FG}
\begin{equation}
\begin{aligned}
y &= (gr)^2 + f_0 + \frac{\xi \ln(gr)}{(gr)^2} +\frac{c_z + c_t
    +c_0}{(gr)^2} + O\left(\frac{\ln r}{r^4}\right)\,,\\
\e^{2T}& = (gr)^2 + a_0 + \frac{\rho \ln(gr)}{(gr)^2} +\frac{c_z}
{(gr)^2} + O\left(\frac{\ln r}{r^4}\right)\,,\\
\e^{2V} &= (gr)^2 + b_0 + \frac{\chi \ln(gr)}{(gr)^2} +\frac{c_t
}{(gr)^2} + O\left(\frac{\ln r}{r^4}\right)\,. 
\label{FG-exp}
\end{aligned}
\end{equation}
Substituting these into the equations \eqref{eqy}, \eqref{eqF} and 
\eqref{eqG'}, yields the coefficients
\begin{equation}
\begin{aligned}
f_0 &=\frac{2k}3\,, \qquad a_0 = b_0 = \frac k2\,,\qquad 
c_0= \frac{(kgq)^2}3\,,\\
\xi &=\frac{k^2}6[1 - 12(gq)^2]\,, \qquad \rho = \chi = \frac{\xi}2\,,
\end{aligned}
\end{equation}
plus the relation
\begin{equation}\label{ctz}
c_z - c_t = \frac{\mu g}2\,.
\end{equation}
Therefore the expansion at large $r$ depends only on the two
constants $c_t$ and $c_z$.\\
We assume in addition that there is an event horizon at $r=r_\text{h}$, and that
the metric functions can be expanded into a Taylor series near 
$r_\text{h}$.\footnote{Note that the latter assumption, while reasonable for
  non-extremal solutions, is violated for the extremal supersymmetric
  black strings found in \cite{Klemm:2000nj}. We will see below how
  the near-horizon expansion looks like in the extremal case.}
Then the Einstein equations imply
\begin{equation}
\begin{aligned}
y &=\frac{12g^2r_\text{h}^4 + 3 k r_\text{h}^2 - 4(kq)^2}{3 r_\text{h}^3}(r - r_\text{h}) \\
&\phantom{{}={}}  -\frac{6 g^2 r_\text{h}^4 + 3 k r_\text{h}^2 - 8(kq)^2 }{3 r_\text{h}^4} (r - r_\text{h})^2 
+ O((r - r_\text{h})^3)\,,\\
\e^{2T} &= a_\text{h} +  \frac{4 a_\text{h} \big[6 g^2 r_\text{h}^4 + (kq)^2\big]}
{r_\text{h}[12g^2r_\text{h}^4 + 3 k r_\text{h}^2 - 4(kq)^2]}(r - r_\text{h})  \\
&\phantom{{}={}}  +\frac{4a_\text{h}[36 g^4 r_\text{h}^8 + 4(kq)^4 - 3 k^3(q r_\text{h})^2
- 6 r_\text{h}^4 (kgq)^2]}{r_\text{h}^2[12g^2r_\text{h}^4 + 3 k r_\text{h}^2 - 4(kq)^2]^2}(r - r_\text{h})^2
 + O((r - r_\text{h})^3)\,,  \\
\e^{2V} &= b_\text{h} (r - r_\text{h}) - \frac{b_\text{h}[6 g^2 r_\text{h}^4 + 3 k
r_\text{h}^2 - 4(kq)^2]}{r_\text{h}[12g^2r_\text{h}^4 + 3 k r_\text{h}^2 - 4(kq)^2]}(r - r_\text{h})^2 
 + O((r - r_\text{h})^3)\,, \label{near-hor-exp}
\end{aligned}
\end{equation}

where
\begin{equation} \label{bh}
b_\text{h} = \frac{12 \mu^2}{a_\text{h} r_\text{h}\big[12g^2r_\text{h}^4 + 3k r_\text{h}^2 - 4(kq)^2\big]}\,.
\end{equation}
Equation \eqref{eqy} actually admits two further expansions for $y$ with different
coefficients. One of them is not compatible with the desired behaviour of $V$
and $T$ close to the horizon, while the other is given by
\eq
y = \alpha_2 (r-r_\text{h})^2 + O((r-r_\text{h})^3)\,, \label{y-extr}
\feq
with
\eq
\alpha_2 = 22g^2 + \frac{5k}{2r_\text{h}^2}\pm\sqrt{228g^4 + \frac{9k^2}{4r_\text{h}^4}
+ \frac{46g^2k}{r_\text{h}^2}}\,. \label{alpha2}
\feq
Notice that the position of the event horizon $r_\text{h}$ is not fixed by
the equations of motion for the expansion \eqref{near-hor-exp}, while
for the second possibility \eqref{y-extr} one finds $f(r_\text{h})=0$, so
$r_\text{h}$ is given in terms of $k$ and $q$. \eqref{y-extr} is related to
the extremal case $\mu=0$ that we shall consider below.

For $q=0$, \eqref{FG-exp} and \eqref{near-hor-exp} reduce correctly to the
expansions at infinity and near the horizon for uncharged black
strings \cite{Mann:2006yi}.

The conditions for a regular event horizon are $y'(r_\text{h}) > 0$ and
$(\e^{2V})'(r_\text{h}) >0$. Using \eqref{near-hor-exp}, these imply $a_\text{h}>0$ and,
in the $k = \pm 1$ case, the further condition
\eq
r_\text{h} > \frac{\sqrt{-18k + 6\sqrt{9 + 192(gq)^2}}}{12 g}\,, \label{bound-rh}
\feq
which gives a minimal value for $r_\text{h}$.

We note that globally regular solutions with $r_\text{h} =0$, which exist for
$q=0$ \cite{Mann:2006yi}, are not allowed for non-vanishing magnetic charge $q$.

From eqns.\ \eqref{eqF} and \eqref{eqG'} it is clear that changing the sign of
$\mu$ leaves $F$ invariant, while $G\to-G$. This means that $V$ and
$T$ are interchanged under $\mu\to-\mu$. Looking at
\eqref{near-hor-exp}, we see that, if a solution with a given value of
$\mu$ describes a black string, then the corresponding solution with
$-\mu$ is a bubble of nothing, which could have been obtained also by a double
analytic continuation of the black string. For $q=0$, such bubble
geometries were considered in \cite{Copsey:2006br}. 

Let us finally determine the near-horizon geometry in the extremal
case $\mu=0$. Expanding \eqref{eqF} near $r=r_\text{h}$ implies then
$f(r_\text{h})=0$. Using \eqref{near-hor-exp}, the terms of order $(r-r_\text{h})^2$
in \eqref{eqF} give the condition $14g^2r_\text{h}^2+k=0$, that, combined
with $f(r_\text{h})=0$, yields $q^2<0$, which is impossible. Thus, for
$\mu=0$, the correct near-horizon expansion for $y$ must be given by
\eqref{y-extr}. Using this in \eqref{eqF'} gives after integration
\eq
\e^F = \e^{2V} = \e^{2T} = C(r-r_\text{h})^{\gamma}(1 + O(r-r_\text{h}))\,, \label{Fmu=0}
\feq
with the exponent
\eq
\gamma = \frac{16g^2r_\text{h}^2 - r_\text{h}^2\alpha_2 + 2k}{r_\text{h}^2\alpha_2}\,,
\feq
and $C$ is an integration constant.
Note that $\gamma$ is positive if and only if we choose the lower sign in
\eqref{alpha2}. In the supersymmetric case $q^2=1/12g^2$, $k=-1$ 
\cite{Klemm:2000nj} one has $r_\text{h}^2 = 1/3g^2$ (cf.\ the following
subsection) and thus $\alpha_2 = 4g^2$ and $\gamma = 3/2$, which is
indeed the correct exponent \cite{Klemm:2000nj}. Introducing the new
coordinate $\rho = (r-r_\text{h})^{\gamma}$, it is easy to see that the
near-horizon geometry of the extremal solutions is AdS$_3$ $\times$
$\cal S$.

\subsection{Exact solutions}
\label{exact-sol}

Despite the complexity of the differential equation \eqref{eqy} it is possible
to find some exact solutions that we list in the following.
\begin{itemize}
\item{$k=0$, $q=0$:}

In this case the equations of motion are solved by
\begin{equation}
\e^{2T} = (gr)^2, \qquad \e^{2V} = y =(gr)^2 -\frac{\mu}{2gr^2}\,.
\end{equation}
Considering $z$ as a coordinate of the transverse space, this can also
be viewed as a black hole, it is the metric found in \cite{Birmingham:1998nr}.

\item{$k$ arbitrary, $q^2 = 1/12g^2$:}

In the case of quantized magnetic charge, one finds two exact
solutions. The first is the supersymmetric magnetic
string \cite{Chamseddine:1999xk,Klemm:2000nj} given by
\eq
\e^{2T} = \e^{2V} = (gr)^{\frac 12}\left(gr + \frac k{3gr}\right)^{\frac
  32}\,, 
\qquad
y = \left(gr + \frac k{3gr}\right)^2\,. \label{susy-sol}
\feq
For $k=-1$ this has an event horizon at $r=r_\text{h}=1/\sqrt 3g$, which is
the limiting value of \eqref{bound-rh}.

The second is a non-extremal black string with finite FG expansion for $y$,
\begin{equation}
\begin{aligned}
y &= (gr)^2 + \frac{2k}3 - \frac{2 k^2}{9(gr)^2}+\frac k{9(gr)^4}\,, \\
\e^{2T}&= \e^{-G} gr\sqrt{(gr)^2 + k}\,,\\
\e^{2V}&= \e^G gr\sqrt{(gr)^2 + k}\,,
\end{aligned}
\end{equation}
where
\begin{equation}
\begin{aligned}
\e^G &= \left[\frac{3(gr)^2(k-2)}{-2+3k(gr)^2-2\sqrt{9(gr)^4-3k(gr)^2+1}}
\right]^{\frac{3g\mu k}2}\nonumber \\
& \qquad\qquad\qquad\qquad
\cdot\left[\frac{5-21k(gr)^2+2\sqrt{13}\sqrt{9(gr)^4-3k(gr)^2+1}}
{((gr)^2+k)(6\sqrt{13}-21k)}\right]^{\frac{3g\mu k}{2\sqrt{13}}}\,,
\end{aligned}
\end{equation}
and
\begin{equation}
\mu = \pm\frac{\sqrt{13}k}{3g}\,. \label{mu}
\end{equation}
Notice that the two signs in \eqref{mu} are related by an interchange
of $T$ and $V$; $\mu\to-\mu$ implies $G\to-G$ and thus $T\to V$, $V\to T$.

For $k=0$ the above metric gives simply AdS$_5$. In the non-trivial
cases $k=\pm 1$ the Kretschmann scalar
$R_{\mu\nu\rho\sigma}R^{\mu\nu\rho\sigma}$ blows up as $r$ goes to
zero. This curvature singularity is naked if $k=1$, while for $k=-1$
it is hidden by an event horizon at $r=r_\text{h}=1/g$ if we choose the lower
sign in \eqref{mu}. At the horizon $\e^{2V}$ vanishes linearly while
$\e^{2T}$ goes to a constant. Comparing with the Fefferman-Graham
expansion \eqref{FG-exp}, we see that this particular black string
solution corresponds to $c_t = -1/8-\sqrt{13}/12$, $c_z = -1/8 +
\sqrt{13}/12$. It has non-vanishing Hawking temperature and
Bekenstein-Hawking entropy. For the upper sign in \eqref{mu} one
obtains a bubble solution similar in spirit to those considered in
\cite{Copsey:2006br}.
\end{itemize}

\subsection{Numerical computation}

So far we have not been able to solve the differential equations \eqref{eq1} to
\eqref{eq4} analytically for general $r_\text{h}$ and $q$.
A numerical evaluation requires some care, since the constants 
$c_t$ and $c_z$ appear as subleading terms in the 
Fefferman-Graham expansion \eqref{FG-exp}.
In subsubsection \ref{scm}
we will show that these constants determine the mass
$M$ and the tension $\cal T$ of the corresponding solution.
In addition, we have to determine the a priori unknown constants $a_\text{h}$ 
and $b_\text{h}$ in the near-horizon expansions \eqref{near-hor-exp}. In 
particular, as we will show in subsection \ref{thermodyn}, 
$a_\text{h}$ is required to compute the area of the event horizon, 
and hence the entropy $S$ and Hawking temperature $T_\text{H}$ of the 
corresponding solution.
With boundary conditions given at $r\eqsim r_\text{h}$, the 
numerical evolution of the solutions to large $r$ therefore has to be 
accurate enough to extract $c_t$ and $c_z$ from the subleading terms in the
asymptotic expansions \eqref{FG-exp}, and also to fix $a_\text{h}$ and 
$b_\text{h}$ a posteriori. 

The boundary conditions taken from the near-horizon expansions in
\eqref{near-hor-exp} depend on the a priori unknown constants
$a_\text{h}$ and $b_\text{h}$. These constants have to be chosen such that the 
asymptotic expansions are of the Fefferman-Graham form as in \eqref{FG-exp}, 
i.e.\ the coefficients in front of $(gr)^2$ should be exactly one.
This is always the case for $y=\e^{-2U}$, but not for $\e^{2T}$ and 
$\e^{2V}$. Since $T$ and $V$ enter the differential equations 
\eqref{eq1} to
\eqref{eq4} only via their first and second derivatives, the respective
solutions are only determined up to an additive constant, i.e.\ the
corresponding warp factors $\e^{2T}$ and $\e^{2V}$ can be rescaled 
by appropriate constants. Such rescalings correspond to rescalings of
respectively $z$ and $t$ in the ansatz for the metric \eqref{ans-metr}.
The freedom of the rescalings is uniquely fixed by the condition that 
the asymptotic expansion has to be of the Fefferman-Graham form, which 
determines $a_\text{h}$ and $b_\text{h}$ in the respective 
near-horizon expansions \eqref{near-hor-exp} of $\e^{2T}$ and $\e^{2V}$.
We can therefore start from arbitrary non-vanishing initial values for 
$a_\text{h}$ and 
$b_\text{h}$, and determine the solution of the differential equations.
The correct values for $a_\text{h}$ and $b_\text{h}$ are then found afterwards 
by dividing our initially chosen values by the corresponding coefficients in 
front of the $(gr)^2$ terms in the asymptotic expansions of our found 
numerical solutions. 

We use \eqref{eq1}, the linear combination 
$\eqref{eq3}-\eqref{eq1}-\eqref{eq2}+\frac{2}{(gr)^2}\eqref{eq4}$
and \eqref{eq4} for the numerical evaluation which we perform with {\tt 
mathematica}.
It turns out that a direct integration 
of the three equations for the three functions 
$y_1=y=\e^{-2U}$,  $y_2=\e^{2T}$, $y_3=\e^{2V}$ does not provide the 
required accuracy: the values for $c_t$ and $c_z$ strongly depend on the 
values of $r\gg r_\text{h}$ which we use to fix the asymptotic expansions.
We therefore first separate the leading near-horizon and leading asymptotic 
behaviours from the unknown subleading contributions by introducing new
functions for which we derive differential equations that can be integrated 
with higher numerical accuracy. 
The original functions $y_i$ are therefore split into a product of two 
functions as\footnote{A similar ansatz has been used in \cite{Copsey:2006br}.}
\begin{equation}\label{yidef}
y_i(u)=\mu_i(u)\bigg(1+\frac{w_i(u)}{1+u^6}\bigg)\,,\qquad i=1,2,3\,,
\end{equation}
where the $\mu_i$ are chosen as follows
\begin{equation}
\mu_i(u)=u^2+f_i-(f_i-g_i+u_\text{h}^2)\frac{1}{1+(u-u_\text{h})^4}
+\bigg(\xi_i\frac{\log u}{u^2}
+\frac{c_i}{u^2}\bigg)\frac{(u-u_\text{h})^4}{1+(u-u_\text{h})^4}\,.
\end{equation}  
We have thereby introduced the dimensionless variables
$u=gr$ and $u_\text{h}=gr_\text{h}$. The constants 
$f_i$, $\xi_i$, $c_i$ 
follow from \eqref{FG-exp} and inherit their names from the expansion of 
$y_1=y$. The constants $g_i$ consider that the
near-horizon expansion of $y_2$ in \eqref{near-hor-exp} starts with 
a constant. We identify
\begin{equation}
f_i=(f_0,a_0,b_0)\,,\qquad
\xi_i=(\xi,\rho,\chi)\,,\qquad
c_i=(c_0,0,0)\,,\qquad
g_i=(0,1,0)\,.
\end{equation}
The functions $\mu_i$, which expand as
\begin{equation}\label{munh}
\mu_{\text{nh},i}(u)
=g_i+2u_\text{h}(u-u_\text{h})+(u-u_\text{h})^2+O((u-u_\text{h})^3)
\end{equation}
near the horizon then have the asymptotic behaviours
\begin{equation}\label{muasy}
\mu_{\text{asy},i}(u)=u^2+f_i+\xi_i\frac{\log u}{u^2}+\frac{c_i}{u^2}
+O\bigg(\frac{1}{u^4}\bigg)\,,
\end{equation}
which are the known parts of the asymptotic expansion of $y_i$
in \eqref{FG-exp} without $c_t$ and $c_z$. 

To match the near-horizon expansion of \eqref{yidef} 
to the near-horizon expansions in \eqref{near-hor-exp}, 
using that $\mu_i$ expands as given in \eqref{munh},
the $w_i$ themselves have to expand as 
\begin{equation}
\begin{aligned}\label{wnh}
w_1(u)&=(1+u_\text{h}^6)\bigg[1+\frac{1}{2u_\text{h}^4}\bigg(u_\text{h}^2k-\frac{4}{3}(kgq)^2\bigg)\bigg]\\
&\phantom{{}={}}
+\bigg[\frac{1+13u_\text{h}^6}{4u_\text{h}^5}
\bigg(u_\text{h}^2k-\frac{4}{3}(kgq)^2\bigg)+6u_\text{h}^5\bigg](u-u_\text{h})
+O((u-u_\text{h})^2)\,,\\
w_2(u)&=(1+u_\text{h}^6)(a_\text{h}-1)\\
&\phantom{{}={}}
-\bigg[\frac{a_\text{h}}{u_\text{h}}\bigg(1+2u_\text{h}^2-5u_\text{h}^6
+2u_\text{h}^8-\frac{3u_\text{h}^2(1+u_\text{h}^6)(12u_\text{h}^2+k)}{12u_\text{h}^4+3u_\text{h}^2k-4(kgq)^2}\bigg)+6u_\text{h}^5\bigg](u-u_\text{h})\\
&\phantom{{}={}}
+O((u-u_\text{h})^2)\,,\\
w_3(u)&=(1+u_\text{h}^6)\bigg[\frac{b_\text{h}}{2u_\text{h}}-1\bigg]\\
&\phantom{{}={}}
+\bigg[\frac{3b_\text{h}}{4u_\text{h}}\frac{1}{12u_\text{h}^4+3u_\text{h}^2k-4(kgq)^2}\bigg(\frac{1+13u_\text{h}^6}{u_\text{h}}\bigg(u_\text{h}^2k-\frac{4}{3}(kgq)^2\bigg)+48u_\text{h}^9\bigg)\\
&\phantom{{}={}+\bigg[}
-6u_\text{h}^5\bigg](u-u_\text{h})
+O((u-u_\text{h})^2)
\,.
\end{aligned}
\end{equation}
This result is used to specify the boundary conditions for integrating 
the differential equations for the $w_i$.
In our numerical evaluation this cannot be done exactly at $u=u_\text{h}$.
We have to specify $w_1(u_0)$, $w_2(u_0)$, $w_3(u_0)$, $w_3'(u_0)$ 
at $u_0=u_\text{h}(1+\varepsilon)$, where the minimal possible value for 
$\varepsilon\ll1$ depends on the numerical integration routine. 
The results presented here are all obtained with
$\varepsilon<10^{-4}$. 
We initially fix $a_\text{h}=b_\text{h}=1$. We then fit to the 
found solutions the two functions $\mu_{\text{asy},i}$ and $\frac{1}{u^2}$.
As explained above, the correct values for $a_\text{h}$ and $b_\text{h}$ 
are then found by dividing their initially chosen values by  
the coefficients of respectively $\mu_{\text{asy},2}$ and
$\mu_{\text{asy},3}$ in the corresponding fit-function. 
To increase the accuracy, it is advantageous 
to repeat this procedure with the new values for $a_\text{h}$ and $b_\text{h}$
until the coefficients in front of $\mu_{\text{nh},2}$,  
$\mu_{\text{nh},3}$ are compatible with one. In this way we can 
achieve an accuracy better than $10^{-8}$ for these coefficients. The final 
fit then yields $c_t$ and $c_z$ as the respective coefficient of the 
second fit function $\frac{1}{u^2}$. A fit of 
the functions $\mu_{\text{asy},1}$ and $\frac{1}{u^2}$ to the result for $y_1$
independently also determines the sum $c_t+c_z$.

\FIGURE[t]{%
\psset{xunit=0.7cm,yunit=0.7cm}
\begin{pspicture}(0,0)(20,20)
\rput[lt](0,20){%
\psset{xunit=5cm,yunit=4cm}
\begin{pspicture}(-0.2,-0.15)(1.125,1.525)
\footnotesize
\psaxes[ticksize=2pt,tickstyle=bottom,Dx=0.2,Dy=0.2]{->}(0,0)(1.02,1.42)
\rput(0,1.475){$a_\text{h}$}
\rput[l](1.05,0){$u_\text{h}$}
\rput(1,1.425){$k=1$}
\rput(1,0.725){$k=-1$}
\readdata{\data}{hahq0.dat}
\dataplot[plotstyle=dots,dotstyle=o,dotsize=1pt]{\data}
\readdata{\data}{hahq0negk.dat}
\dataplot[plotstyle=dots,dotstyle=o,dotsize=1pt]{\data}
\psdots[plotstyle=dots,dotstyle=o,dotsize=1pt](0.1,1.3)(0.1125,1.3)(0.125,1.3)(0.1375,1.3)(0.15,1.3)(0.1625,1.3)(0.175,1.3)(0.1875,1.3)(0.2,1.3)
\rput[l](0.25,1.3){$gq=0$}
\readdata{\data}{hahq0625.dat}
\psset{linestyle=dashed,dash=4pt 4pt,plotstyle=curve}
\dataplot{\data}
\readdata{\data}{hahq0625negk.dat}
\dataplot{\data}
\psline(0.1,1.2)(0.2,1.2)
\rput[l](0.25,1.2){$gq=0.0625$}
\readdata{\data}{hahq1.dat}
\psset{linestyle=dashed,dash=2pt 2pt,plotstyle=curve}
\dataplot{\data}
\readdata{\data}{hahq1negk.dat}
\dataplot{\data}
\psline(0.1,1.1)(0.2,1.1)
\rput[l](0.25,1.1){$gq=0.1$}
\readdata{\data}{hahq135860.dat}
\psset{linestyle=solid}
\dataplot{\data}
\readdata{\data}{hahq135860negk.dat}
\dataplot{\data}
\psline(0.1,1)(0.2,1)
\rput[l](0.25,1){$gq=0.13586$}
\readdata{\data}{hahq2886.dat}
\dataplot[linestyle=dotted,linewidth=0.5pt]{\data}
\readdata{\data}{hahq2886negk.dat}
\dataplot[linestyle=dotted,linewidth=0.5pt]{\data}
\psline[linestyle=dotted,linewidth=0.5pt](0.1,0.9)(0.2,0.9)
\rput[l](0.25,0.9){$gq=\frac{1}{\sqrt{12}}$}
\end{pspicture}
}
\rput[rt](20,20){%
\psset{xunit=5cm,yunit=0.705263cm}
\begin{pspicture}(-0.2,-0.75)(1.125,8.75)
\footnotesize
\psaxes[ticksize=2pt,tickstyle=bottom,Dx=0.2,Dy=1]{->}(0,0)(1.02,8.1)
\rput(0,8.5){$b_\text{h}$}
\rput[l](1.05,0){$u_\text{h}$}
\rput(1,5.1){$k=1$}
\rput(1,3.5){$k=-1$}
\readdata{\data}{hbhq0.dat}
\dataplot[plotstyle=dots,dotstyle=o,dotsize=1pt]{\data}
\readdata{\data}{hbhq0negk.dat}
\dataplot[plotstyle=dots,dotstyle=o,dotsize=1pt]{\data}
%
\readdata{\data}{hbhq0625.dat}
\psset{linestyle=dashed,dash=4pt 4pt,plotstyle=curve}
\dataplot{\data}
\readdata{\data}{hbhq0625negk.dat}
\dataplot{\data}
%
\readdata{\data}{hbhq1.dat}
\psset{linestyle=dashed,dash=2pt 2pt,plotstyle=curve}
\dataplot{\data}
\readdata{\data}{hbhq1negk.dat}
\dataplot{\data}
%
\readdata{\data}{hbhq135860.dat}
\psset{linestyle=solid}
\dataplot{\data}
\readdata{\data}{hbhq135860negk.dat}
\dataplot{\data}
\psline(0.04,5.4)(0.0525,5.4)
%
\readdata{\data}{hbhq2886.dat}
\dataplot[linestyle=dotted,linewidth=0.5pt]{\data}
\readdata{\data}{hbhq2886negk.dat}
\dataplot[linestyle=dotted,linewidth=0.5pt]{\data}
\end{pspicture}
}
\rput[b](10,0){%
\psset{xunit=11.0416666cm,yunit=10.72cm}
\begin{pspicture}(-0.1,-0.4)(1.1,0.225)
\footnotesize
\psaxes[ticksize=2pt,tickstyle=bottom,Dx=0.2,Dy=0.1]{->}(0,0)(0,0.201)(1.02,-0.41)
\rput(0.5,-0.35){$c_t$}
\rput(0.7,-0.125){$c_z$}
\rput(0.8,-0.125){$c_t$}
\rput(0.9,0.175){$c_z$}
\rput[l](1.05,0){$u_\text{h}$}
\rput(0.5,-0.15){$k=1$}
\rput(0.5,0.1){$k=-1$}
\readdata{\data}{hctq0.dat}
\dataplot[plotstyle=dots,dotstyle=o,dotsize=1pt]{\data}
\readdata{\data}{hczq0.dat}
\dataplot[plotstyle=dots,dotstyle=o,dotsize=1pt]{\data}
\readdata{\data}{hctq0negk.dat}
\dataplot[plotstyle=dots,dotstyle=o,dotsize=1pt]{\data}
\readdata{\data}{hczq0negk.dat}
\dataplot[plotstyle=dots,dotstyle=o,dotsize=1pt]{\data}
%
%
\readdata{\data}{hctq0625.dat}
\psset{linestyle=dashed,dash=4pt 4pt,plotstyle=curve}
\dataplot{\data}
\readdata{\data}{hczq0625.dat}
\dataplot{\data}
\readdata{\data}{hctq0625negk.dat}
\dataplot{\data}
\readdata{\data}{hczq0625negk.dat}
\dataplot{\data}
\psline(0.04,4.6)(0.0525,4.6)
\rput[l](0.0575,4.6){$gq=0.0625$}
\readdata{\data}{hctq1.dat}
\psset{linestyle=dashed,dash=2pt 2pt,plotstyle=curve}
\dataplot{\data}
\readdata{\data}{hczq1.dat}
\dataplot{\data}
\readdata{\data}{hctq1negk.dat}
\dataplot{\data}
\readdata{\data}{hczq1negk.dat}
\dataplot{\data}
\psline(0.04,5)(0.0525,5)
\rput[l](0.0575,5){$gq=0.1$}
\readdata{\data}{hctq135860.dat}
\psset{linestyle=solid}
\dataplot{\data}
\readdata{\data}{hczq135860.dat}
\dataplot{\data}
\readdata{\data}{hctq135860negk.dat}
\dataplot{\data}
\readdata{\data}{hczq135860negk.dat}
\dataplot{\data}
\psline(0.04,5.4)(0.0525,5.4)
\rput[l](0.0575,5.4){$gq=0.13586$}
%
\readdata{\data}{hctq2886.dat}
\dataplot[linestyle=dotted,linewidth=0.5pt]{\data}
\readdata{\data}{hczq2886.dat}
\dataplot[linestyle=dotted,linewidth=0.5pt]{\data}
\readdata{\data}{hctq2886negk.dat}
\dataplot[linestyle=dotted,linewidth=0.5pt]{\data}
\readdata{\data}{hczq2886negk.dat}
\dataplot[linestyle=dotted,linewidth=0.5pt]{\data}
\psline[linestyle=dotted,linewidth=0.5pt](0.04,5.8)(0.0525,5.8)
\rput[l](0.0575,5.8){$gq=\frac{1}{\sqrt{12}}$}
\end{pspicture}
}
\end{pspicture}
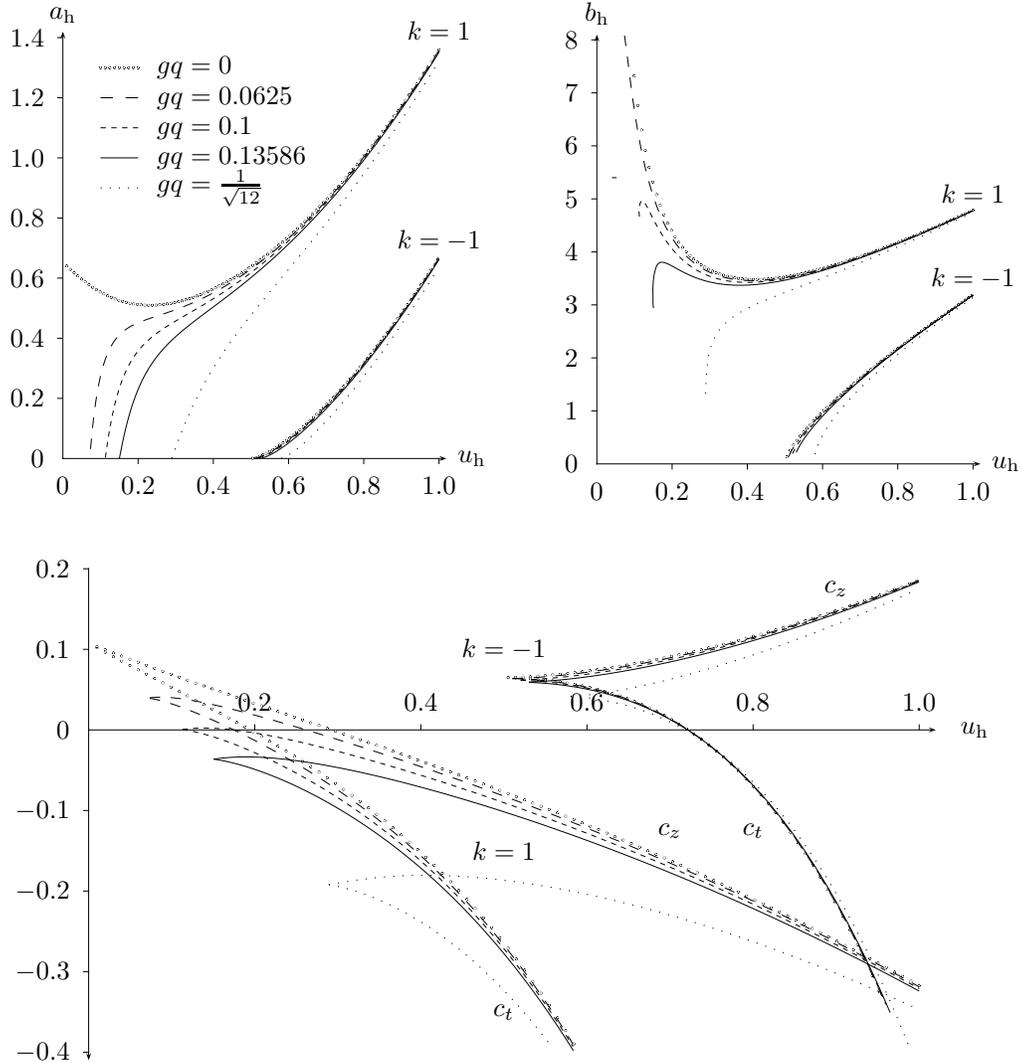
\caption{$a_\text{h}$, $b_\text{h}$, $c_t$, $c_z$ as functions of 
$u_\text{h}$ for $k=1$ and $k=-1$ at five values of the magnetic charge 
$q$.\label{hahbhctcz}}}

The information on $c_t$ and $c_z$ is directly encoded in 
the asymptotic expansion of the second factor in \eqref{yidef}. 
It is designed in such a way that
the second derivatives $\frac{1}{2}w''_i$ at 
large $u$ directly give $c_t+c_z$, $c_t$, $c_z$. However, in practice 
only  $\frac{1}{2}w''_1$ becomes constant at large $u$ such that 
one can directly read off $c_t+c_z$. In the other two cases,
a fit of the 
three functions $u^2$, $u^4$, $u^6$ to the solution based on the 
correct values for $a_\text{h}$ and $b_\text{h}$
yields $c_t$ and $c_z$ as the 
coefficients of $u^2$.
We also find a very small admixture of $u^4$ and $u^6$ terms 
spoils the constancy of the derivatives 
$\frac{1}{2}w''_2$, $\frac{1}{2}w''_3$ at large $u$.
To obtain accurate values for $c_t$ and $c_z$, the fits to $w_i$ require
a determination of the solutions to much larger values of $u$ than 
required for a fit to the $y_i$.
 
The numerical results for the functions $w_i$ are nevertheless useful to
find the regimes in which we can trust the numerical results and to 
determine which interval of large $u$ should be used for the fits to
$y_i$ to achieve the highest accuracy. 
We find that our 
ansatz \eqref{yidef} works fine at least for $0<u_\text{h}\lesssim 2$, which 
is sufficient for our purposes. For larger $u_\text{h}$ it is enough
to slightly modify \eqref{yidef} by replacing $u\to\frac{u}{u_\text{h}}$ and  
$(u-h)\to(\frac{u}{u_\text{h}}-1)$ in the interpolating 
factors. Furthermore, in the regime $0<u_\text{h}\lesssim 2$ a restriction to
$u_\text{h}<u\lesssim 100$ avoids the regime of increasing noise 
above $u\eqsim120$, and it suffices to fit the asymptotics with high 
precision. This we have checked by reproducing the
exact results of subsection \ref{exact-sol}. 
For fits in the interval $0.8\,u_\text{max}\le u\le u_\text{max}$ with 
$u_\text{max}=100$ we obtain the highest relative accuracy, which in any case 
is better than $10^{-5}$ for $c_t$ and $c_z$.

For given $g=1$, $k=\pm1$ and charge $q$ we compute 
$a_\text{h}(u_\text{h})$, $b_\text{h}(u_\text{h})$, $c_t(u_\text{h})$,
$c_z(u_\text{h})$ for sufficiently many values of $u_\text{h}$.
The corresponding results for  $u_\text{min}\le u_\text{h}\le 1$ are
presented in figure \ref{hahbhctcz} for $k=1$ and $k=-1$ at 
five values of the magnetic charge $q$. 
These results allow us to determine
the inverse Hawking temperature $T_\text{H}$ as a function of the entropy 
$S$, and the free energy $F$ as a function of $T_\text{H}$.
To determine the critical charge $q_\text{crit}$, we fit a linear function
to data points which we concentrate around the estimated turning point 
and vary $q$ until the slope of the fit function is compatible with zero. This 
allows us to determine $gq_\text{crit}=0.13586(1)$. We have determined the 
error, which only affects the last digit, by the lower and upper bound 
for $q_\text{crit}$
which undoubtedly are below and above $q_\text{crit}$, respectively. 
At $q_\text{crit}$ we then fit a cubic polynomial to the data points around 
the estimated turning point. 
This then fixes the turning point more precisely to 
$(S\frac{Gg^2}{LV_k},\frac{g}{T_\text{H}})
=(0.017278(\tiny\begin{matrix}-0.000006 \\ +0.000010\end{matrix}),7.12673(\tiny\begin{matrix}-0.00013 \\ +0.00007\end{matrix}))$, where we estimate the
errors from the corresponding fits in which $q$ assumes the value of 
the lower or upper bound of $q_\text{crit}$.
Finally, we should remark that the corresponding data-files 
for all the 
plots are available as parts of the source files of this paper. 
They are called {\tt [xquantity][yquantity]q[value of q as integer][neg/pos]k.dat}.

\section{Properties of magnetic black strings}\label{properties}

\subsection{Conserved quantities}\label{conserved}

\subsubsection{Standard counterterm method}
\label{scm}

To compute the mass and tension, which we expect to be encoded in the constants
$c_t$ and $c_z$ appearing in \eqref{FG-exp}, we use the counterterm procedure
for spacetimes with negative cosmological constant proposed
in \cite{Balasubramanian:1999re}.
One obtains a finite quasilocal stress tensor
\begin{align}\label{stresstensor}
T^{ij}  =  \frac 2{\sqrt{-\gamma}}\frac{\delta I}{\delta\gamma_{ij}}
\end{align}
by adding to the action \eqref{action} a counterterm $I_{ct}$,
\begin{equation} \label{Iren}
I_{\mathrm{ren}} = I_0 + I_{\mathrm{ct}}(\gamma_{ij})\,,
\end{equation}
where $\gamma_{ij}$ denotes the induced metric on the boundary 
$\partial{\cal M}$, which we take to be a hypersurface at constant
radial coordinate $r$. By requiring cancellation of divergences in the
limit $r\to\infty$, one finds the explicit expression for
$I_{\mathrm{ct}}$ \cite{Balasubramanian:1999re}, 
\begin{align}
I_{\mathrm{ct}}&=-\frac 1{8\pi G}\int_{\mathcal{\partial M}}\de^4 x \,
\sqrt{-\gamma}\left\{-3g\left(1 + \frac{\cal R}{12g^2}\right)\quad  
\right.\nonumber \\
&\phantom{{}={}-\frac 1{8\pi G}\int_{\mathcal{\partial M}}\de^4 x \,
\sqrt{-\gamma}\left\{\right.}
+\left.\frac{\ln(gr)}{2g}\left[\frac 1{4g^2}\left(\frac 13{\cal R}^2 -
{\cal R}_{ij}{\cal R}^{ij}\right) + F_{ij}F^{ij} \right]\right\}\,, \label{Ict}
\end{align}
where $\cal R$ and ${\cal R}^{ij}$ are the curvature and the Ricci tensor
associated with the metric $\gamma$. The first term in the second line
of \eqref{Ict} is the usual expression required to cancel logarithmic
divergences \cite{Skenderis:2000in} that appear in odd dimensions,
while the second one is due to the presence of additional matter
fields (in our case a U(1) gauge field) in the bulk
\cite{TaylorRobinson:2000xw}. A logarithmic contribution to the
counterterms also appears naturally in a reformulation of the
holographic renormalization procedure in terms of the extrinsic
curvature \cite{Papadimitriou:2004ap}.

Varying the renormalized action \eqref{Iren} with respect to the
boundary metric $\gamma_{ij}$ leads to the holographic energy-momentum tensor
\begin{equation}
\begin{aligned}\label{stress}
T^{ij}
&=\frac 2{\sqrt{-\gamma}}\frac{\delta I_{\mathrm{ren}}}{\delta\gamma_{ij}}\\ 
&= \frac 1{8\pi G}\left\{K^{ij} - K\gamma^{ij} + \frac 1{2g}G^{ij}\right. \\
&\phantom{{}={}\frac 1{8\pi G}\left\{\right.} 
+ \frac{\ln(gr)}{2g}\left[\frac 1{4g^2}\left(\frac 13\gamma^{ij}{\cal R}^2 -
\gamma^{ij}{\cal R}_{kl}{\cal R}^{kl} - \frac 43{\cal R}{\cal R}^{ij} 
+ 4{\cal R}^{ikjl}{\cal R}_{kl}\right.\right.\\
&\left.\left.
\qquad\qquad\qquad\qquad\qquad\quad
+2\square\lp{\mathcal R}^{ij}-\frac12\ga^{ij}{\mathcal R}\rp
+\frac23\lp\ga^{ij}\square-\nabla^i\nabla^j\rp{\mathcal R}
\right)\right. \\
&\phantom{{}={}\frac 1{8\pi G}\left\{+ \frac{\ln(gr)}{2g}\left[\right.\right.} 
+ \gamma^{ij}F_{kl}F^{kl} - 4 F^{ik}F^{jl}\gamma_{kl}
\bigg]\bigg\}\,.
\end{aligned}
\end{equation}
Let $^4{\cal S}_t$ be a spacelike hypersurface at constant $t$, with
unit normal $n$, and 
$\Sigma = {^4{\cal S}}_t\,\cap\,\partial{\cal M}$, with induced metric 
$\sigma$. Then, for any Killing vector field $\xi$ associated with an isometry
of the boundary four-metric, one defines the conserved charge
\begin{equation} \label{Q}
Q_{\xi} = \int_{\Sigma} \de^3 x \sqrt{\sigma}\,n^{i}\,T_{ij}\,\xi^{j}\,.
\end{equation}
The charge associated to time translation invariance 
($\xi = \partial_t$) is the mass $M$ of the spacetime. Evaluating
\eqref{Q} for $r\to\infty$ we obtain
\eq
M = \frac {LV_k}{16\pi Gg}\left[c_z - 3c_t + \frac{k^2}{12}\right]\,, \label{M}
\feq
where $L$ denotes the period of the $z$ direction, and $V_k$ is the area
of the angular sector\footnote{$V_k$ is finite if one compactifies the 
two-surface $\cal S$ to a torus ($k=0$) or to a Riemann surface of
genus $h>1$ ($k=-1$). One has then $V_0 = |{\mathrm{Im}}\tau|$, with
$\tau$ the Teichm\"uller parameter of the torus, and, 
using Gauss-Bonnet, $V_{-1} = 4\pi(h-1)$. In the case of
non-compact $\cal S$ one can define a mass per unit volume.},
$V_k = \int\de\theta\de\varphi\,S(\theta)$.

A naive application of \eqref{Q} to compute the conserved charge associated to
$\xi = \partial_z$ yields zero, because there is no momentum along the
string\footnote{Solutions with momentum along $z$ will be discussed
in section \ref{waves}.}. Nevertheless, there exists a non-vanishing charge
corresponding to translations in $z$, namely the string tension
$\cal T$ \cite{Traschen:2001pb,Townsend:2001rg,Traschen:2003jm,Harmark:2003dg},
which can also be computed using \eqref{Q}, but now $n$ is the unit normal
to a surface of constant $z$. In our case we get for the tension per unit time
\eq
{\cal T} = \frac{V_k}{16\pi Gg}\left[3c_z -  c_t -
  \frac{k^2}{12}\right]\,. 
\label{T}
\feq
Note that \eqref{M} and \eqref{T} coincide with the results obtained in
\cite{Mann:2006yi} for the uncharged case.

The vacuum is given by the supersymmetric solution \eqref{susy-sol},
which corresponds to $c_t = c_z = k^2/24$ and has thus  vanishing mass
and tension. Therefore, the standard regularization does not produce
any vacuum energy. Notice, however, that this procedure suffers from
an ambiguity. In fact, to the minimal counterterm action \eqref{Ict}
one can always add terms quadratic in the Riemann tensor, Ricci tensor
and Ricci scalar of the boundary. As in four dimensions the variation
of the Euler term
\begin{equation}
\mathcal{E}_4 = \sqrt{-\gamma}\left({\cal R}_{ijkl}{\cal R}^{ijkl} 
- 4{\cal R}_{ij}{\cal R}^{ij} + {\cal R}^2\right)
\end{equation}
vanishes, the general quadratic term that produces the ambiguity simplifies to
\eq \label{deltaIct}
\Delta I_{\mathrm{ct}} = -\frac 1{8\pi G g^3}\, \int_{\partial{\cal M}}
\de^4 x \, \sqrt{-\gamma} \left(\alpha {\cal R}_{ij}{\cal R}^{ij} 
+ \beta {\cal R}^2 \right)\,,
\feq
with $\alpha$ and $\beta$ denoting arbitrary constants.
This yields an additional contribution to the stress tensor,
\eq
\begin{aligned}
&\Delta T^{ij} = \frac 1{8\pi G g^3}\left[\gamma^{ij}\left(\alpha {\cal R}_{kl}
{\cal R}^{kl} + \beta {\cal R}^2\right) 
- 4 \left(\alpha {\cal R}^{ikjl}{\cal R}_{kl}
+ \beta {\cal R}{\cal R}^{ij}\right)\vphantom{\frac12}\right.\nonumber\\
&\qquad\qquad\qquad\quad\left.-2\alpha\square\left({\mathcal R}^{ij}-\frac12\gamma^{ij}{\mathcal R}\rp
-(\alpha+2\beta)\lp\ga^{ij}\square-\nabla^i\nabla^j\right){\mathcal R}\right]\,;
\label{deltaT}
\end{aligned}
\feq
in particular, the variations of the energy \eqref{M} and the tension \eqref{T} are
\eq
\Delta M = - \frac{k^2 L V_k}{4\pi Gg}\, \left(\alpha + 2 \beta\right)\qquad
\mbox{and} \quad\Delta {\cal T} =  \frac{k^2 V_k}{4\pi Gg}\,
\left(\alpha + 2 \beta
\right)\,. \label{DeltaMT}
\feq
Therefore, conserved quantities are well determined only with the
further specification of the prescription that is assumed regarding
quadratic terms in the counterterm action.

\subsubsection{Holographic stress tensor and conformal anomaly}
\label{anomaly}

An important information on the CFT dual to the black string solutions
\eqref{ans-metr} is encoded in the expectation value of its energy-momentum
tensor, that we wish to compute now. The metric of the background upon which
the dual field theory resides is found by rescaling
\begin{equation}
h_{ij} = \lim_{r\to\infty}\frac 1{(gr)^2}\,\gamma_{ij}\,,
\end{equation}
which yields
\begin{equation}
h_{ij}\,\de x^i \de x^j = -\de t^2 + \de z^2 + \frac 1{g^2}\de\Omega_k^2\,,
\end{equation}
and so the conformal boundary, where the $\mathcal{N} = 4$ SU$(N)$ 
$\textrm{SYM}$ theory lives, is $\bR \times \text{S}^1 \times {\cal S}$.

The stress tensor expectation value $\langle {\hat T}_{jk}\rangle$ can
be computed using the relation\cite{Myers:1999ps}
\begin{equation}
\sqrt{-h}\, h^{ij}\langle{\hat T}_{jk}\rangle = \lim_{r\to\infty}
\sqrt{-\gamma}\,\gamma^{ij}\,T_{jk}\,,
\end{equation}
which gives
\begin{equation}
\begin{aligned}
\langle {\hat T}^t_{\phantom{t}t} \rangle 
&= \frac g{16\pi G}\left[3 c_t - c_z - \frac{k^2}{12}
\right]\,, \qquad 
\langle {\hat T}^z_{\phantom{z}z} \rangle 
= \frac g{16\pi G}\left[3 c_z - c_t - \frac{k^2}{12}\right]\,, \\
\langle {\hat T}^{\theta}_{\phantom{\theta}\theta} \rangle 
&= \langle {\hat T}^{\varphi}_{\phantom{\varphi}
\varphi} \rangle = -\frac g{16\pi G}\left[c_t + c_z + \left(k g q\right)^2 -
\frac{k^2}6\right]\,,
\end{aligned}
\end{equation}
i.e.\ an anisotropic perfect fluid form.
As expected, this stress tensor is not traceless,
\begin{equation}\label{traccia}
\langle {\hat T}^i_{\phantom{i}i}\rangle 
= \frac {gk^2}{96\pi G}\left[1 - 12\left(g q\right)^2
\right]\,.
\end{equation}
The first part of \eqref{traccia} matches exactly the conformal anomaly of the
boundary CFT coming from the background curvature
\cite{Henningson:1998gx,deHaro:2000xn},
\eq
\mathcal{A} = \frac{N^2}{32\pi^2}\left({\cal R}_{ij}{\cal R}^{ij} - \frac 13
{\cal R}^2\right)\,,
\feq
if we use the AdS/CFT dictionary $N^2=\pi/(2Gg^3)$. The second part of
\eqref{traccia}, proportional to $q^2$, results from the coupling of
the CFT to a background gauge field. The two contributions exactly
cancel when the magnetic charge assumes the value
\begin{align}\label{quantizz}
q^2 = \frac 1{12g^2}\,.
\end{align}
This is precisely the behaviour we expect; in fact choosing
\eqref{quantizz} in \eqref{FG-exp} has the effect of cancelling the
logarithmic terms in the Fefferman-Graham expansion, which produce
the anomaly. Notice that the generalized Dirac quantization condition
\eqref{quantizz} was found in \cite{Chamseddine:1999xk,Klemm:2000nj}
by requiring supersymmetry. Maldacena and Nu\~nez showed that it
corresponds to a twisting of the dual SYM theory
\cite{Maldacena:2000mw}: Putting a supersymmetric field theory on a
curved manifold generally breaks supersymmetry, because one will not
have a Killing spinor obeying $(\partial_i + \omega_i)\,\epsilon = 0$,
where $\omega_i$ denotes the spin connection. If, however, the field
theory has a global R-symmetry, it can be coupled to an external gauge
field that couples to the R-symmetry current. If we choose this
external gauge field to be equal to the spin connection, $A_i =
\omega_i$, we can find a covariantly constant spinor since
$(\partial_i + \omega_i - A_i)\,\epsilon = \partial_i\epsilon$, which
vanishes for constant $\epsilon$. The resulting theory is called
`twisted', because the coupling to the external gauge field
effectively changes the spins of all fields. The requirement $A_i =
\omega_i$ yields precisely the charge quantization condition \eqref{quantizz}.

Notice finally that the ambiguity due to the quadratic contributions
\eqref{deltaIct} to the action does not affect the Weyl anomaly, because
\begin{equation}
\langle\Delta {\hat T}^i_{\;\;i} \rangle = 0\,,
\end{equation}
as it can be seen from \eqref{deltaT}.

\subsubsection{Kounterterm procedure}
\label{kounterterms}

A different approach to regularize both the conserved quantities and
the Euclidean action for asymptotically AdS spacetimes is given by
the Kounterterm proposal \cite{Olea:2005gb,Olea:2006vd}, considering covariant
boundary terms depending on both extrinsic and intrinsic quantities,
instead of the Gibbons-Hawking term plus intrinsic counterterms,
\begin{equation}
I_{\mathrm{ren}}=\frac 1{16\pi G}\int_{\mathcal M}\de^{d+1} x \sqrt{-g} \left(
 R - 2 \Lambda \right) - c_d\int_{\partial{\cal M}}
\de^{d}xB_d(\gamma ,\mathcal{R}(\gamma ),K)\,. \label{IrenKT}
\end{equation}
$B_d$ is a polynomial in the induced metric, the boundary Riemann
tensor and the extrinsic curvature. With a suitable choice of $B_d$
one is able to achieve a well-posed variational principle and to
solve the regularization problem at the same time. The main
advantage of this procedure is that it provides a closed formula for
the charges in all dimensions. This is a consequence of the use of
geometrical boundary terms related to topological invariants and
Chern-Simons forms which are not obtained through the algorithm
given by holographic renormalization. Therefore, to obtain 
their explicit form, we do not face the technical difficulties of
standard AdS gravity regularization.

We now focus on the main features of the procedure and specialize to
the five-dimensional case. The corresponding coupling constant $c_4$
is fixed demanding the total action to be stationary under arbitrary
on-shell variations of the fields that respect the asymptotic form
of the metric in asymptotically AdS spacetimes, i.e.,
\begin{equation}
R_{\mu \nu }^{\alpha \beta }+ g^{2}\delta _{\lbrack \mu
\nu]}^{[\alpha\beta]} = 0 \label{ALAdS}
\end{equation}
and
\begin{equation}
K_{j}^{i}=g\delta _{j}^{i}\,. \label{K=delta}
\end{equation}
As it has been argued in \cite{Papadimitriou:2004ap}, a Dirichlet
boundary condition on the metric $\gamma _{ij}$ does not really make
sense in spacetimes with conformal boundary, as it is the case of
asymptotically 
AdS spaces. Indeed, the boundary metric blows up as the boundary is
reached. This can be seen as a motivation to introduce the regular
asymptotic condition \eqref{K=delta}, because it does not induce
additional divergences in the variation of the action, and yet it is
compatible with the idea of holographic reconstruction of the
spacetime.

The boundary term in five dimensions is given by
\begin{equation*}
B_{4}=\sqrt{-\gamma }\delta _{\lbrack
j_{1}j_{2}j_{3}]}^{[i_{1}i_{2}i_{3}]}K_{i_{1}}^{j_{1}}
\left(\mathcal{R}_{i_{2}i_{3}}^{j_{2}j_{3}}-
K_{i_{2}}^{j_{2}}K_{i_{3}}^{j_{3}}+ \frac{g^{2}}{3}\delta
_{i_{2}}^{j_{2}} \delta _{i_{3}}^{j_{3}}\right)\,,
\end{equation*}
and is multiplied by a coupling constant $c_4 = 1/(128\pi Gg^2)$.

Conserved quantities are defined as Noether charges associated to
asymptotic Killing vectors $\xi $. Their expression appears
naturally split into two parts,
\begin{equation}
Q(\xi )=q(\xi )+q_{0}(\xi )\,, \label{QKT}
\end{equation}
where the quantity
\begin{equation}
\begin{aligned}\label{qxi}
q(\xi ) &=\int_{\Sigma }\de^{3}x \sqrt{\sigma }\,\left(u_{j}\,
q_{i}^{j}\,\xi
^{i}\right)\,,  \\
q_{i}^{j} &=\frac{1}{64\pi Gg^{2}}\delta _{\lbrack
i_{1}i_{2}\dots i_{4}]}^{[jj_{2}\dots j_{4}]}K_{i}^{i_{1}}\delta
_{j_{2}}^{i_{2}} \left(R_{j_{3}j_{4}}^{i_{3}i_{4}}+g^{2}\delta
_{\lbrack j_{3}j_{4}]}^{[i_{3}i_{4}]}\right)\,,
\end{aligned}
\end{equation}
provides, in general, the mass and the angular momentum for
point-like solutions, but also for topological black holes. It can
be noticed that the above formula is identically vanishing for the
AdS vacuum. Therefore, the second contribution in \eqref{QKT}
\begin{equation}
\begin{aligned}\label{qijfact}
q_{0}(\xi ) &=\int_{\Sigma }\de^{3}x \,\sqrt{\sigma }
\,\left(u_{j}\, q_{(0)i}^{j}\,\xi ^{i}\right)\,,   \\
q_{(0)i}^{j} &=-\frac{1}{128\pi Gg^{2}}\,\delta _{\lbrack
i_{1}i_{2}\dots i_{4}]}^{[jj_{2}\dots j_{4}]}\, \left(\delta
_{j_{2}}^{i_{2}}\!K_{i}^{i_{1}}\!+\!\delta
_{i}^{i_{2}}\!K_{j_{2}}^{i_{1}}\!\right)
\left(\mathcal{R}_{j_{3}j_{4}}^{i_{3}i_{4}}
-K_{j_{3}}^{i_{3}}K_{j_{4}}^{i_{4}}+g^{2}
\delta _{j_{3}}^{i_{3}}\delta _{j_{4}}^{i_{4}}\right)\,,
\end{aligned}
\end{equation}
is truly a covariant formula for the vacuum energy for any
asymptotically AdS spacetime 
(see \cite{Cheng:2005wk} for the vacuum energy results obtained in the standard regularization approach).

For the quantized charge case \eqref{quantizz}, evaluating the first
piece $q(\xi )$ in the conserved charges formula \eqref{QKT}, we
obtain the same values as in the Dirichlet regularization shown
above, for the energy \eqref{M} if $\xi =\partial _t$ and the
tension \eqref{T} if $\xi =\partial_z$. For the second part, $
q_0(\xi)$, one gets
\begin{equation}
\begin{aligned}\label{VE}
q_{0}(\partial_t) &=-\frac 1{16\pi Gg}\frac{k^2}{24}LV_{k}\,,\qquad
q_{0}(\partial_z) &=\frac 1{16\pi Gg}\frac{k^2}{24}V_{k}\,,
\end{aligned}
\end{equation}
such that there is an additional contribution to the total mass and
tension in the Kounterterm formalism. One could eventually match
these results with the ones obtained using intrinsic counterterms
through the addition of quadratic terms in the curvature
\eqref{deltaIct}. In fact, the ambiguity pointed out in
\eqref{DeltaMT} is able to reproduce the same vacuum energy and
tension for supersymmetric magnetic strings if $\alpha$ and $\beta$
satisfy
\begin{align*}
\alpha + 2 \beta = \frac 1{96}\,,
\end{align*}
because there is no an a priori reasoning to rule out the existence
of a vacuum energy for a supersymmetric solution in asymptotically
AdS gravity.

One might also supplement the Kounterterms with the
logarithmic terms in \eqref{Ict}, which would allow the
regularization of the conserved charges in the general case.

\subsection{Thermodynamics}
\label{thermodyn}

The Hawking temperature of the black string solutions is obtained by
requiring the absence of conical singularities in the Euclidean section of the
metric \eqref{ans-metr}. Setting $t=i\tau$ and using \eqref{near-hor-exp},
the $(\tau,r)$-part of the near-horizon metric becomes
\begin{equation} \label{cone}
\de\sigma^2 = \frac{b_\text{h} \, \alpha}{4}\, \rho^2 \de\tau^2 + \de\rho^2,
\end{equation}
where
\begin{equation}
\alpha = y'(r_\text{h}) = \frac{12g^2r_\text{h}^4 + 3 k r_\text{h}^2 - 4(kq)^2}{3r_\text{h}^3}\,,
\end{equation}
and the new radial coordinate $\rho$ is defined by
\begin{equation}
\de\rho = \frac{\de r}{\sqrt{\alpha}\sqrt{r-r_\text{h}}}\,.
\end{equation}
\eqref{cone} is smooth at $\rho=0$ if $\tau$ is identified modulo
$4\pi/\sqrt{b_\text{h}\alpha}$. This gives the Hawking temperature
\begin{equation} \label{TH}
T_\text{H} 
= \frac 1{4\pi}\sqrt{b_\text{h}\alpha} 
= \frac{|\mu |}{2\pi\sqrt{a_\text{h}}\,r_\text{h}^2}\,.
\end{equation}
In the last step, we used \eqref{bh}.

%
\begin{figure}
\begin{center}
\subfigure[$k=1$]{\label{STinvposk}%
\setlength{\unit}{0.03125\textwidth}
\psset{xunit=42cm,yunit=1.2cm}
\begin{pspicture}(-0.012,3.5)(0.15,10.75)
\footnotesize
\psaxes[ticksize=2pt,tickstyle=bottom,Dx=0.02,Dy=1,Oy=4]{->}(0,4)(0.122,10.1)
\rput(0,10.375){$\frac{g}{T_\text{H}}$}
\rput[l](0.126,4){$S/\frac{LV_k}{Gg^2}$}
\readdata{\data}{STinvq0.dat}
\dataplot[plotstyle=dots,dotstyle=o,dotsize=1pt]{\data}
\psdots[plotstyle=dots,dotstyle=o,dotsize=1pt](0.0415625,4.5)(0.043125,4.5)
(0.0448675,4.5)(0.04625,4.5)(0.0478125,4.5)(0.049375,4.5)(0.0509375,4.5)(0.0525,4.5)
\rput[l](0.0575,4.5){$gq=0$}
\readdata{\data}{STinvq0625.dat}
\psset{linestyle=dashed,dash=4pt 4pt,plotstyle=curve}
\dataplot{\data}
\psline(0.04,4.85)(0.0525,4.85)
\rput[l](0.0575,4.85){$gq=0.0625$}
\readdata{\data}{STinvq1.dat}
\psset{linestyle=dashed,dash=2pt 2pt,plotstyle=curve}
\dataplot{\data}
\psline(0.04,5.2)(0.0525,5.2)
\rput[l](0.0575,5.2){$gq=0.1$}
\readdata{\data}{STinvq135860.dat}
\psset{linestyle=solid}
\dataplot{\data}
\psline(0.04,5.55)(0.0525,5.55)
\rput[l](0.0575,5.55){$gq=0.13586$}
\psframe(0.0164,7.12677)(0.0181,7.12672)
\readdata{\data}{STinvq2886.dat}
\dataplot[linestyle=dotted,linewidth=0.5pt]{\data}
\psline[linestyle=dotted,linewidth=0.5pt](0.04,5.9)(0.0525,5.9)
\rput[l](0.0575,5.9){$gq=\frac{1}{\sqrt{12}}$}
\psline[linecolor=mygray](0.01728,7.12673)(0.0425,10)
\psline[linecolor=mygray](0.01728,7.12673)(0.1525,8.95)
\rput*[lt](0.0425,10){%
\begin{pspicture}(-0.06,-0.625)(0.05,0.425)
\psset{xunit=0.2cm,yunit=0.2cm}
\tiny\mygray
\psset{linecolor=mygray}
\psaxes[ticksize=2pt,ticks=none,labels=none,linestyle=dashed,dash=2pt 2pt,linewidth=0.125pt]{-}(0,0)(-8,-2)(8,2)
\psframe(-8,-2)(8,2)
\psline(-8,-2)(-8,-2.25)\rput(-8,-3.25){$0.01648$}
\psline(0,-2)(0,-2.25)\rput(0,-3.25){$0.01728$}
\psline(8,-2)(8,-2.25)\rput(8,-3.25){$0.01808$}
\psline(-8,-2)(-8.25,-2)\rput[r](-8.5,-2){$7.12671$}
\psline(-8,0)(-8.25,0)\rput[r](-8.5,0){$7.12673$}
\psline(-8,2)(-8.25,2)\rput[r](-8.5,2){$7.12675$}
\readdata{\data}{STinvq135860zoom.dat}
\dataplot[plotstyle=dots,dotsize=1pt,showpoints=true]{\data}
\end{pspicture}
}
\end{pspicture}
}
\subfigure[$k=-1$]
{\label{STinvnegk}%
\psset{xunit=11.34cm,yunit=1.2cm}
\begin{pspicture}(0.1,3.5)(0.7,10.75)
\footnotesize
\psaxes[ticksize=2pt,tickstyle=bottom,Dx=0.1,Dy=1,Oy=4,labels=y]{->}(0.15,4)(0.60,10.1)
\rput[t](0.15,3.8){$0.15$}
\rput[t](0.25,3.8){$0.25$}
\rput[t](0.35,3.8){$0.35$}
\rput[t](0.45,3.8){$0.45$}
\rput[t](0.55,3.8){$0.55$}
\rput(0.15,10.375){$\frac{g}{T_\text{H}}$}
\rput[l](0.6125,4){$S/\frac{LV_k}{Gg^2}$}
\readdata{\data}{STinvq0negk.dat}
\dataplot[plotstyle=dots,dotstyle=o,dotsize=1pt]{\data}
\psdots[plotstyle=dots,dotstyle=o,dotsize=1pt](0.35,7.6)(0.3625,7.6)(0.375,7.6)(0.3875,7.6)(0.4,7.6)
\rput[l](0.425,7.6){$gq=0$}
\readdata{\data}{STinvq0625negk.dat}
\psset{linestyle=dashed,dash=4pt 4pt,plotstyle=curve}
\dataplot{\data}
\psline(0.35,8)(0.4,8)
\rput[l](0.425,8){$gq=0.0625$}
\readdata{\data}{STinvq1negk.dat}
\psset{linestyle=dashed,dash=2pt 2pt,plotstyle=curve}
\dataplot{\data}
\psline(0.35,8.4)(0.4,8.4)
\rput[l](0.425,8.4){$gq=0.1$}
\readdata{\data}{STinvq135860negk.dat}
\psset{linestyle=solid}
\dataplot{\data}
\psline(0.35,8.8)(0.4,8.8)
\rput[l](0.425,8.8){$gq=0.13586$}
\psframe(0.0164,7.12677)(0.0181,7.12672)
\readdata{\data}{STinvq1negk.dat}
\psset{linestyle=dashed,dash=2pt 2pt,plotstyle=curve}
\psset{linestyle=dotted,linewidth=0.75pt,plotstyle=curve}
\dataplot[linestyle=dotted,linewidth=0.5pt]{\data}
\psline[linestyle=dotted,linewidth=0.5pt](0.35,9.2)(0.4,9.2)
\rput[l](0.425,9.2){$gq=\frac{1}{\sqrt{12}}$}
\end{pspicture}
}
\caption{The inverse Hawking temperature $\frac{1}{T_\text{H}}$
as a function of the entropy $S$ \subref{STinvposk} for $k=1$ and 
 \subref{STinvnegk} for $k=-1$
at five values of
the magnetic charge $q$.
The critical charge for a first order phase transition present in 
\subref{STinvposk} is given by $q_\text{crit}=0.13586(1)$. 
In the first two graphs of \subref{STinvposk} with 
$q<q_\text{crit}$ one finds two local extrema and a turning point.
At $q=q_\text{crit}$ the extrema and turning point merge into a saddle point,
which vanishes at $q>q_\text{crit}$.
The graphs of \subref{STinvnegk} nearly coincide for the five values $q$ 
in the plotted range.
\label{STinv}}
\end{center}
\end{figure}
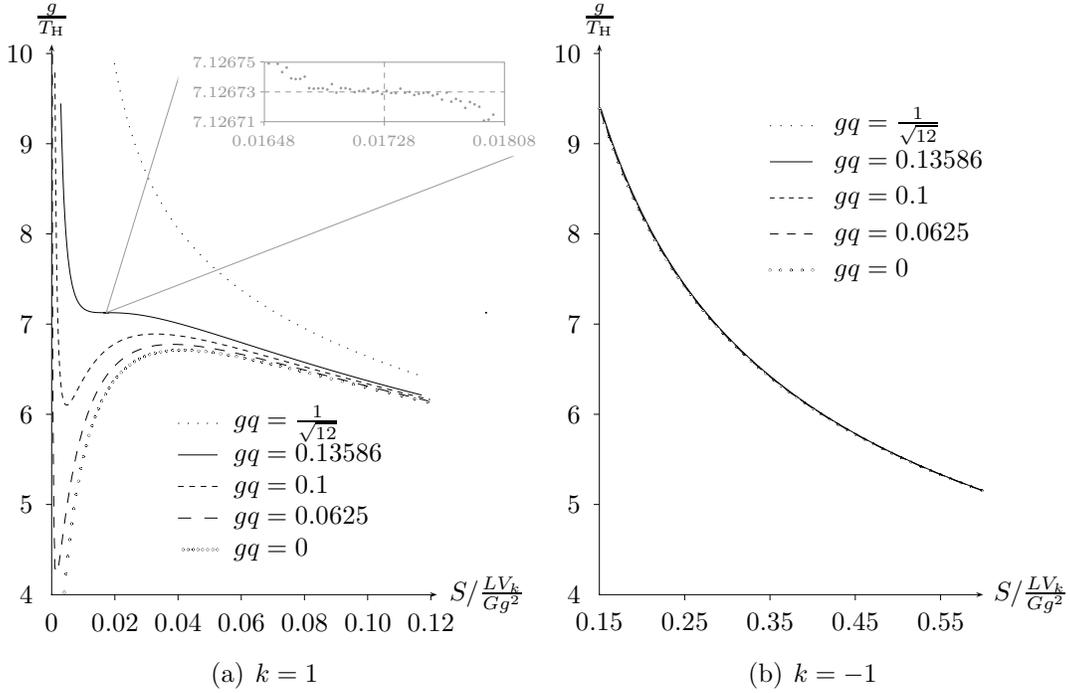

\FIGURE[t]{%
\setlength{\unit}{0.03125\textwidth}
\psset{xunit=20cm,yunit=2.5cm}
\begin{pspicture}(0,-1.125)(0.775,1.125)
\rput(0.15,0){%
\begin{pspicture}(0.06,-1.125)(0.35,1.125)
\scriptsize
\psaxes[ticksize=2pt,tickstyle=bottom,Dx=0.05,Dy=0.4,labels=none]{->}(0.1,-0.8)(0.24,0.84)
\rput(0.235,-0.9125){$\frac{T_\text{H}}{g}$}
\rput(0.1,1.0125){$F/\frac{LV_k}{100Gg}$}
\rput[t](0.1,-0.875){$0.1$}
\rput[t](0.15,-0.875){$0.15$}
\rput[t](0.2,-0.875){$0.2$}
\rput[r](0.0925,-0.8){$-0.8$}
\rput[r](0.0925,-0.4){$-0.4$}
\rput[r](0.0925,0){$0$}
\rput[r](0.09,0.4){$0.4$}
\rput[r](0.0925,0.8){$0.8$}
\readdata{\data}{TFq0625.dat}
\psset{linestyle=solid,dash=4pt 4pt,plotstyle=curve}
\dataplot{\data}
\rput(0.125,-0.075){$\olett{1}$}
\rput(0.19,0.11){$\olett{2}$}
\rput(0.16,-0.4){$\olett{3}$}
\rput(0.17,0.6){$gq=0.0625$}
\end{pspicture}
}
\rput(0.345,0){%
\begin{pspicture}(0.06,-1.125)(0.35,1.125)
\scriptsize
\psaxes[ticksize=2pt,tickstyle=bottom,Dx=0.05,Dy=0.4,labels=none]{->}(0.1,-0.8)(0.24,0.84)
\rput(0.235,-0.9125){$\frac{T_\text{H}}{g}$}
\rput(0.1,1.0125){$F/\frac{LV_k}{100Gg}$}
\rput[t](0.1,-0.875){$0.1$}
\rput[t](0.15,-0.875){$0.15$}
\rput[t](0.2,-0.875){$0.2$}
\rput[r](0.0925,-0.8){$-0.8$}
\rput[r](0.0925,-0.4){$-0.4$}
\rput[r](0.0925,0){$0$}
\rput[r](0.09,0.4){$0.4$}
\rput[r](0.0925,0.8){$0.8$}
\readdata{\data}{TFq1.dat}
\psset{linestyle=solid,dash=4pt 4pt,plotstyle=curve}
\dataplot{\data}
\rput(0.17,0.6){$gq=0.1$}
\end{pspicture}
}
\rput(0.54,0){%
\begin{pspicture}(0.06,-1.125)(0.35,1.125)
\scriptsize
\psaxes[ticksize=2pt,tickstyle=bottom,Dx=0.05,Dy=0.4,labels=none]{->}(0.1,-0.8)(0.24,0.84)
\rput(0.235,-0.9125){$\frac{T_\text{H}}{g}$}
\rput(0.1,1.0125){$F/\frac{LV_k}{100Gg}$}
\rput[t](0.1,-0.875){$0.1$}
\rput[t](0.15,-0.875){$0.15$}
\rput[t](0.2,-0.875){$0.2$}
\rput[r](0.0925,-0.8){$-0.8$}
\rput[r](0.0925,-0.4){$-0.4$}
\rput[r](0.0925,0){$0$}
\rput[r](0.09,0.4){$0.4$}
\rput[r](0.0925,0.8){$0.8$}
\readdata{\data}{TFq135860.dat}
\psset{linestyle=solid,dash=4pt 4pt,plotstyle=curve}
\dataplot{\data}
\rput(0.17,0.6){$gq=0.13586$}
\end{pspicture}
}
\rput(0.735,0){%
\begin{pspicture}(0.06,-1.125)(0.35,1.125)
\scriptsize
\psaxes[ticksize=2pt,tickstyle=bottom,Dx=0.05,Dy=0.4,labels=none]{->}(0.1,-0.8)(0.24,0.84)
\rput(0.235,-0.9125){$\frac{T_\text{H}}{g}$}
\rput(0.1,1.0125){$F/\frac{LV_k}{100Gg}$}
\rput[t](0.1,-0.875){$0.1$}
\rput[t](0.15,-0.875){$0.15$}
\rput[t](0.2,-0.875){$0.2$}
\rput[r](0.0925,-0.8){$-0.8$}
\rput[r](0.0925,-0.4){$-0.4$}
\rput[r](0.0925,0){$0$}
\rput[r](0.09,0.4){$0.4$}
\rput[r](0.0925,0.8){$0.8$}
\readdata{\data}{TFq2886.dat}
\psset{linestyle=solid,dash=4pt 4pt,plotstyle=curve}
\dataplot{\data}
\rput(0.17,0.6){$gq=\frac{1}{\sqrt{12}}$}
\end{pspicture}
}
\end{pspicture}
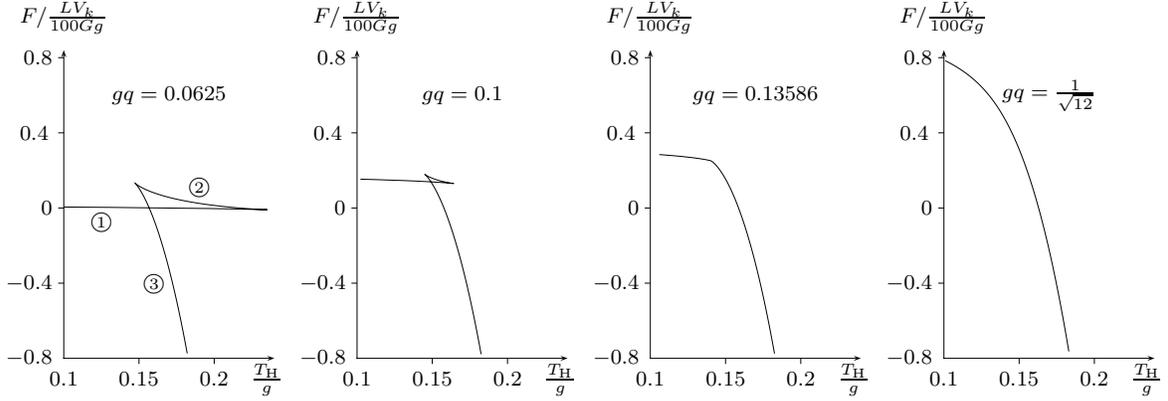
\caption{The free energy $F$ as a function of the Hawking temperature 
$T_\text{H}$ for $k=1$ at four values of the magnetic charge $q$. 
The critical charge
is given by $q_\text{crit}=0.13586(1)$. In the first two diagrams with 
$q<q_\text{crit}$ one finds three branches. At $q=q_\text{crit}$ the 
branches merge into a single one, which remains also at $q>q_\text{crit}$.}
\label{phaseposk}}

Computation of the area of the event horizon
\begin{equation}\label{Ahor}
A = \int_{r=r_\text{h}} \e^T r_\text{h}^2\, S(\theta) \de z\,
\textrm{d}\theta \de\varphi
\end{equation}
yields for the entropy
\begin{equation}\label{S}
S 
= \frac A{4G} 
= \frac{\sqrt{a_\text{h}}\,r_\text{h}^2\, L \, V_k}{4 G}\,.
\end{equation}
Note that for $\mu\ge 0$ the thermodynamic quantities \eqref{M}, \eqref{T},
\eqref{TH} and \eqref{S} obey the Smarr-type
formula\footnote{Note that a Smarr-type formula has been shown to hold in full generality \cite{Papadimitriou:2005ii}.}
\begin{equation}\label{smarr}
 M + \mathcal{T} L = T_\text{H} S\,.
\end{equation}
As a confirmation of the results obtained so far, we compute the
Euclidean action
\begin{equation}
I_{\mathrm E} 
= -\frac 1{4\pi G}\int\sqrt g\left[\frac R4 + 3g^2 - \frac 14 F_{\mu\nu}
      F^{\mu\nu}
+ \frac 1{6\sqrt 3}\epsilon^{\mu\nu\alpha\beta\gamma} F_{\mu\nu}
      F_{\alpha\beta}A_{\gamma}\right]\de^5x + I_{\mathrm{GH}}
      + I_{\mathrm{ct}}\,,
\end{equation}
where
\begin{equation}
I_{\mathrm{GH}} = \frac 1{8\pi G}\int\de^4x \sqrt{\gamma}K\,,
\end{equation}
and $I_{\mathrm{ct}}$ is given by \eqref{Ict}, with $\gamma$ instead
of $-\gamma$, but with the same overall sign. In our case, the
Chern-Simons term does not contribute, and $I_{\text E}$ reduces on-shell to
\eq
I_{\mathrm E} = \frac 1{4\pi G}\int\sqrt g\left[2g^2 + \frac 16 F_{\mu\nu}
      F^{\mu\nu}\right]\textrm{d}^5x + I_{\mathrm{GH}} + I_{\mathrm{ct}}\,.
\feq
Plugging in our expressions for $\sqrt g$ and $F^2$, the bulk term reads
\eq
I_{\mathrm{E},\,\mathrm{bulk}} 
= \frac{\beta L V_k}{4\pi G}\int_{r_\text{h}}^R \e^{F+U}r^2
\left[2g^2 + \frac{k^2q^2}{3r^4}\right]\textrm{d}r\,,
\feq
where $\beta$ denotes the inverse temperature and $R$ is some cutoff to be sent
to infinity at the end of the calculation. Using the equation of
motion \eqref{eq1}, this can be integrated to give
\eq \label{I_Ebulk}
I_{\mathrm{E},\,\mathrm{bulk}} = \frac{\beta L V_k}{8\pi G}\left[V'\e^{F-U}r^2
\right]_{r_\text{h}}^R\,.
\feq
Evaluation of the boundary terms yields
\begin{equation}
I_{\mathrm{GH}} + I_{\mathrm{ct}} 
= -\frac{\beta L V_k}{8\pi G}\e^F\left[\e^{-U}r^2
\left(F' + \frac 2r\right) - 3gr^2 - \frac k{2g} 
+ \frac{k^2\ln(gr)}{gr^2}\left(q^2
- \frac 1{12g^2}\right)\right]_{r=R}\,.
\end{equation}
\FIGURE[H]{%
\psset{xunit=20cm,yunit=0.57143cm}
\begin{pspicture}(0.06,0)(0.25,-9.84375)
\scriptsize
\psaxes[ticksize=2pt,tickstyle=bottom,Dx=0.05,Dy=1,labels=none]{->}(0.1,-1)(0.24,-8.2)
\rput(0.235,-0.55){$\frac{T_\text{H}}{g}$}
\rput[t](0.1,-8.5){$F/\frac{LV_k}{100Gg}$}
\rput[b](0.1,-0.75){$0.1$}
\rput[b](0.15,-0.75){$0.15$}
\rput[b](0.2,-0.75){$0.2$}
\rput[r](0.0925,-1){$-1$}
\rput[r](0.0925,-2){$-2$}
\rput[r](0.0925,-3){$-3$}
\rput[r](0.0925,-4){$-4$}
\rput[r](0.0925,-5){$-5$}
\rput[r](0.0925,-6){$-6$}
\rput[r](0.09,-7){$-7$}
\rput[r](0.0925,-8){$-8$}
\readdata{\data}{TFq0625negk.dat}
\psset{linestyle=solid,dash=4pt 4pt,plotstyle=curve}
\dataplot{\data}
\end{pspicture}
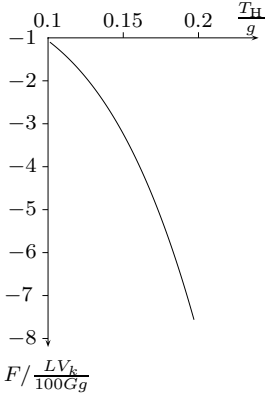
\caption{The free energy $F$ as a function of the Hawking temperature 
$T_\text{H}$ for $k=-1$ at $gq=0.13586(1)$.\label{phasenegk}}}
Adding this to \eqref{I_Ebulk} and taking the limit $R\to\infty$, the
final result takes the form
\eq
I_{\mathrm E} = \beta(M - T_\text{H} S)\,,
\feq
with $M$, $T_\text{H}$ and $S$ given by \eqref{M}, \eqref{TH} and \eqref{S}
respectively. The Helmholtz free energy is thus
\eq
F = \frac{I_{\mathrm E}}{\beta} = M - T_\text{H} S\,,
\feq
which correctly coincides with the result we would have obtained by simply
Legendre transforming the mass. Using \eqref{smarr}, one gets
\eq
F = -\mathcal{T}L\,. \label{free-energy}
\feq

Note that the Smarr formula \eqref{smarr} is a simple consequence
of the scaling behaviour
\begin{equation}
F(T_\text{H}, \lambda L, q) = \lambda F(T_\text{H}, L, q)\,.
\end{equation}
Deriving this with respect to $\lambda$, using
$\mathcal{T} = -(\partial F/\partial L)_{T_\text{H}, q}$ and setting 
$\lambda=1$ gives \eqref{free-energy} and thus \eqref{smarr}.

Figures \ref{STinvposk} and \ref{STinvnegk} show the
inverse temperature $\beta=1/T_\text{H}$ as a function of the entropy
$S$ (in units of $LV_k/G$) for $k=1$ and $k=-1$ respectively. In both
cases, the curves, which  were obtained numerically, are drawn
for five different values of the magnetic charge $q$. In
the $k=-1$ case the graphs nearly coincide for the chosen
values of $q$. Note that the functional
relationship $\beta=\beta(S)$ represents one of the equations of state. We see 
that for $k=-1$ the entropy decreases monotonically with $\beta$ for all
values of $q$, so the heat capacity
\begin{equation}
c_{L,q} = T_\text{H}\left(\frac{\partial S}{\partial T_\text{H}}\right)_{L,q}
\end{equation}
is always positive, which is a necessary condition for local thermodynamic
stability. 
This situation changes drastically for $k=1$: If the magnetic charge is smaller
than the critical value $gq_{\mathrm{crit}}=0.13586(1)$, there are three
branches of black string solutions, with the middle branch being
thermodynamically unstable. 
The left one is present only for non-vanishing $q$,
because otherwise no extremal $k=1$ black strings exist. At
$q=q_{\mathrm{crit}}$ the turning points of $\beta(S)$ merge, and
disappear for $q>q_{\mathrm{crit}}$, so that we are left with one single
branch of stable black strings. 
Notice that the limit $S\to 0$ corresponds to
the extremal solutions that have zero Hawking temperature, and horizon
coordinate $r_\text{h}$ given by the limiting value of \eqref{bound-rh}. 
The $\beta(S)$ curve reminds us of the $P(V)$ van der Waals equation of state,
where the pressure $P$ is replaced here by $\beta$ and the volume $V$ by
$S$. This analogy was noticed for the first time and explored in detail in
\cite{Chamblin:1999tk,Chamblin:1999hg} for the case of electrically charged
AdS black holes\footnote{Rotation was included
in \cite{Caldarelli:1999xj}.}. We will have to say more about this later.

\FIGURE[b]{%
\setlength{\unit}{0.03125\textwidth}
\psset{xunit=0.93848276cm,yunit=1.7755102cm}
\begin{pspicture}(3.5,-0.4)(10.75,4.5)
\footnotesize
\psaxes[ticksize=2pt,tickstyle=bottom,Dx=1,Dy=1,Ox=4]{->}(4,0)(10.1,4)
\rput(10.5,0){$\frac{g}{T_\text{H}}$}
\rput[l](3.5,4.25){$M/\frac{LV_k}{100Gg}$}
\rput(5.5,4){$k=1$}
\rput(8.5,4){$k=-1$}
\scriptsize
\readdata{\data}{TinvMq0.dat}
\dataplot[plotstyle=dots,dotstyle=o,dotsize=1pt]{\data}
\readdata{\data}{TinvMq0negk.dat}
\dataplot[plotstyle=dots,dotstyle=o,dotsize=1pt]{\data}
\psdots[plotstyle=dots,dotstyle=o,dotsize=1pt](4.25,0.7)
(4.375,0.7)
(4.5,0.7)
(4.625,0.7)
(4.75,0.7)
\rput[l](4.875,0.7){$gq=0$}
\readdata{\data}{TinvMq0625.dat}
\psset{linestyle=dashed,dash=4pt 4pt,plotstyle=curve}
\dataplot{\data}
\readdata{\data}{TinvMq0625negk.dat}
\dataplot{\data}
\psline(4.25,0.9)(4.75,0.9)
\rput[l](4.875,0.9){$gq=0.0625$}
\readdata{\data}{TinvMq1.dat}
\psset{linestyle=dashed,dash=2pt 2pt,plotstyle=curve}
\dataplot{\data}
\readdata{\data}{TinvMq1negk.dat}
\dataplot{\data}
\psline(4.25,1.1)(4.75,1.1)
\rput[l](4.875,1.1){$gq=0.1$}
\readdata{\data}{TinvMq135860.dat}
\psset{linestyle=solid}
\dataplot{\data}
\readdata{\data}{TinvMq135860negk.dat}
\dataplot{\data}
\psline(4.25,1.3)(4.75,1.3)
\rput[l](4.875,1.3){$gq=0.13586$}
\readdata{\data}{TinvMq2886.dat}
\dataplot[linestyle=dotted,linewidth=0.5pt]{\data}
\readdata{\data}{TinvMq2886negk.dat}
\dataplot[linestyle=dotted,linewidth=0.5pt]{\data}
\psline[linestyle=dotted,linewidth=0.5pt](4.25,1.5)(4.74,1.5)
\rput[l](4.875,1.5){$gq=\frac{1}{\sqrt{12}}$}
\end{pspicture}
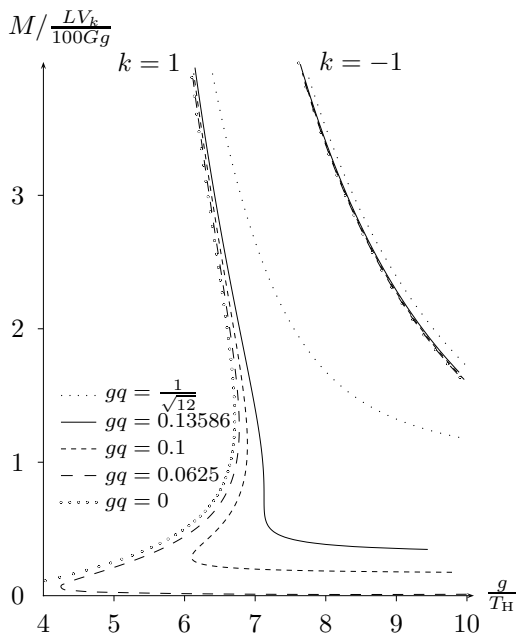
\caption{The mass $M$ in dependence of the inverse Hawking temperature 
$T_\text{H}$ for $k=1$ and $k=-1$ at five values of the magnetic charge $q$.
\label{TinvMq}}}

The free energy (in units of $LV_k/100Gg$) as a function of temperature is
shown in figure \ref{phaseposk} ($k=1$) for four different values of the
magnetic charge $q$ and in figure \ref{phasenegk} ($k=-1$), for a single 
value
of $q$.

In the hyperbolic case $k=-1$, there is always one
single branch. For $k=1$ and small charge $q<q_{\mathrm{crit}}$, starting at
the left of the plot (low temperature), there is a single branch of free
energy, corresponding to stable small black strings, which we shall refer to
as branch 1. At a certain temperature, branches 2 (unstable black strings) and
3 (stable large black strings) appear and separate from each other at higher
temperatures. At some still larger temperature, branches 1 and 2 meet and
disappear, whereas branch 3 continues to the right.

If we raise the magnetic charge, the swallowtail (i.e.\ the triangle
confined by the three branches) becomes smaller and finally disappears at
$q=q_{\mathrm{crit}}$. At this critical point, the first order phase
transition degenerates and becomes of higher order. Finally, for
$q>q_{\mathrm{crit}}$, only a single stable branch remains.

Notice that the free energy does not go to zero for $T_\text{H}\to 0$, which
means that
in general the extremal black string has non-vanishing $F$. It is easy to see
that $F=0$ for the supersymmetric solution \eqref{susy-sol}, which therefore
represents some sort of background. 
\FIGURE[t]{%
\setlength{\unit}{0.03125\textwidth}
\psset{xunit=5.4166666cm,yunit=0.725cm}
\begin{pspicture}(-0.1,-1)(1.1,11)
\footnotesize
\psaxes[ticksize=2pt,tickstyle=bottom,Dx=0.2,Dy=2]{->}(0,0)(1,10.1)
\rput(0,10.5){$\frac{g}{T_\text{H}}$}
\rput(1.05,0){$gr_\text{h}$}
%
\readdata{\data}{hTinvq0.dat}
\dataplot[plotstyle=dots,dotstyle=o,dotsize=1pt]{\data}
\psdots[plotstyle=dots,dotstyle=o,dotsize=1pt](0.3,1)(0.325,1)(0.35,1)(0.375,1)(0.4,1)
\rput[l](0.42,1){$gq=0$}
\readdata{\data}{hTinvq0625.dat}
\psset{linestyle=dashed,dash=4pt 4pt,plotstyle=curve}
\dataplot{\data}
\psline(0.3,1.6)(0.4,1.6)
\rput[l](0.42,1.6){$gq=0.0625$}
\readdata{\data}{hTinvq1.dat}
\psset{linestyle=dashed,dash=2pt 2pt,plotstyle=curve}
\dataplot{\data}
\psline(0.3,2.2)(0.4,2.2)
\rput[l](0.42,2.2){$gq=0.1$}
\readdata{\data}{hTinvq135860.dat}
\psset{linestyle=solid}
\dataplot{\data}
\psline(0.3,2.8)(0.4,2.8)
\rput[l](0.42,2.8){$gq=0.13586$}
\readdata{\data}{hTinvq2886.dat}
\dataplot[linestyle=dotted,linewidth=0.5pt]{\data}
\psline[linestyle=dotted,linewidth=0.5pt](0.3,3.4)(0.4,3.4)
\rput[l](0.42,3.4){$gq=\frac{1}{\sqrt{12}}$}
\end{pspicture}
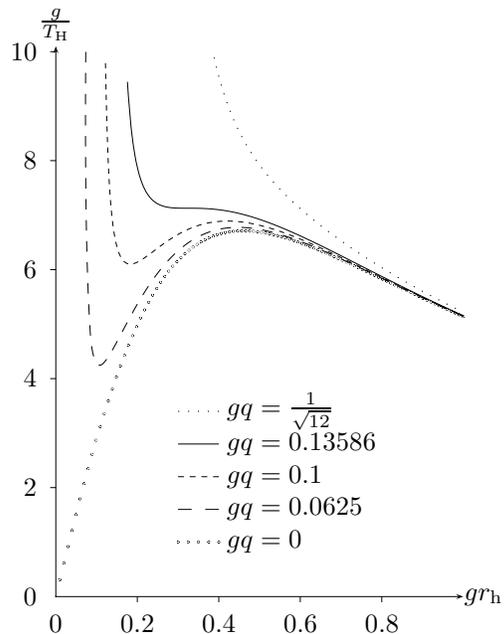
\caption{The inverse Hawking temperature $\frac{1}{T_\text{H}}$ 
in dependence of the horizon $r_\text{h}$  
for $k=1$ at five values of the magnetic charge $q$.
\label{STinvq}}
}

We are however free to add finite
counterterms to \eqref{Ict}, e.g.\ a term proportional to
\eq
\int\textrm{d}^4x \sqrt{-\gamma}F_{ij}F^{ij}\,. \label{delta_I_F2}
\feq
This would give an additional contribution to the Euclidean action that depends
on the magnetic charge. By adjusting the prefactor of \eqref{delta_I_F2} one
can probably obtain a free energy that vanishes for the extremal solution,
although we did not check this explicitly.

Finally, by performing the double analytic continuation $\chi= it$, $\tau= iz$, corresponding to a change of the sign of $\mu$, the black string solutions become static magnetically charged bubble of nothing solutions. The $\text{S}^1$ factor of the metric pinches of at the radius $r_\text{h}$, and the solution is regular if the spatial coordinate $\chi$ is identified modulo $s=1/T_\text{H}=4\pi/\sqrt{b_\text{h}\alpha}$.
These are zero-temperature solutions, in the $L\rightarrow\infty$ limit, with the same spatial asymptotic structure ${\mathcal S}\times \text{S}^1$ as the black strings. As shown in figure~\ref{STinvq}, the length of the $\text{S}^1$ is fixed by the size $r_\text{h}$ of the bubble. Note that in presence of a magnetic charge there is a minimal size of the bubble, unlike in the uncharged $k=1$ case studied in \cite{Copsey:2006br}. Also, these uncharged $k=1$ bubbles of nothing where shown to exist only under some critical length of the $\text{S}^1$. If, however, one adds magnetic charge, bubble solutions exist for any length of the $\text{S}^1$. The ground state is given by the lowest total energy bubble, which is given (in terms of the black string quantities) by
\begin{equation}
E_\text{b}=-\frac{\mathcal T}{T_\text{H}}=\frac{F}{LT_\text{H}}.
\end{equation}
The energy of the bubble is plotted in figure~\ref{TinvEposk} as a function of the size $s$ of the asymptotic S$^1$. For non-vanishing magnetic charge below the critical value there are three branches of bubbles, that merge in a single branch above the critical charge. Therefore, there is a first order phase transition between a small size bubble and a large size one for charges lower than the critical charge $q_{\mathrm{crit}}$, that disappears for higher charge. Therefore, the quantum phase transition that occurs in the strongly coupled dual gauge theory as one varies the size of the S$^1$ found in \cite{Copsey:2006br} for the uncharged $k=1$ bubbles becomes a quantum phase transition between the states dual to the small/large bubbles, and then disappears for $q>q_{\mathrm{crit}}$.
\FIGURE[t]{%
\setlength{\unit}{0.03125\textwidth}
\psset{xunit=0.44705882cm,yunit=1.25cm}
\begin{pspicture}(0,-2.5)(34,2.5)
\rput(4.25,0){%
\begin{pspicture}(2.5,-1.5)(11,3.5)
\scriptsize
\psaxes[ticksize=2pt,tickstyle=bottom,Dx=2,Dy=1,Ox=4]{->}(4,0)(4,-1)(10.2,3.05)
\rput(10.5,0.2){$gs$}
\rput[l](3,3.25){$E_\text{b}/\frac{V_k}{100Gg^2}$}
\readdata{\data}{TinvEq0625.dat}
\psset{linestyle=solid,dash=4pt 4pt,plotstyle=curve}
\dataplot{\data}
\rput[l](4.75,2.5){$gq=0.0625$}
\end{pspicture}
}
\rput(12.75,0){%
\begin{pspicture}(2.5,-1.5)(11,3.5)
\scriptsize
\psaxes[ticksize=2pt,tickstyle=bottom,Dx=2,Dy=1,Ox=4]{->}(4,0)(4,-1)(10.2,3.05)
\rput(10.5,0.2){$gs$}
\rput[l](3,3.25){$E_\text{b}/\frac{V_k}{100Gg^2}$}
\readdata{\data}{TinvEq1.dat}
\psset{linestyle=solid,dash=4pt 4pt,plotstyle=curve}
\dataplot{\data}
\rput[l](4.75,2.5){$gq=0.1$}
\end{pspicture}
}
\rput(21.25,0){%
\begin{pspicture}(2.5,-1.5)(11,3.5)
\scriptsize
\psaxes[ticksize=2pt,tickstyle=bottom,Dx=2,Dy=1,Ox=4]{->}(4,0)(4,-1)(10.2,3.05)
\rput(10.5,0.2){$gs$}
\rput[l](3,3.25){$E_\text{b}/\frac{V_k}{100Gg^2}$}
\readdata{\data}{TinvEq135860.dat}
\psset{linestyle=solid,dash=4pt 4pt,plotstyle=curve}
\dataplot{\data}
\rput[l](4.75,2.5){$gq=0.13586$}
\end{pspicture}
}
\rput(29.75,0){%
\begin{pspicture}(2.5,-1.5)(11,3.5)
\scriptsize
\psaxes[ticksize=2pt,tickstyle=bottom,Dx=2,Dy=1,Ox=4]{->}(4,0)(4,-1)(10.2,3.05)
\rput(10.5,0.2){$gs$}
\rput[l](3,3.25){$E_\text{b}/\frac{V_k}{100Gg^2}$}
\readdata{\data}{TinvEq2886.dat}
\psset{linestyle=solid,dash=4pt 4pt,plotstyle=curve}
\dataplot{\data}
\rput[l](7.25,2.5){$gq=\frac{1}{\sqrt{2}}$}
\end{pspicture}
}
\end{pspicture}
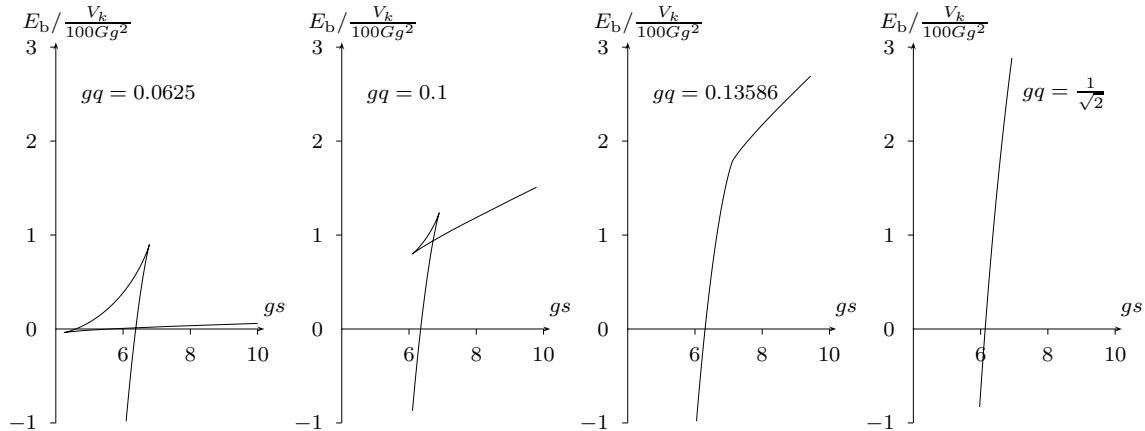
\caption{The energy $E_\text{b}$ as a function of the of the size $s=\frac{1}{T_\text{H}}$ of the asymptotic S$^1$ for $k=1$ at four values of the magnetic charge $q$. The critical charge
is given by $q_\text{crit}=0.13586(1)$. In the first two diagrams with 
$q<q_\text{crit}$ one finds three branches. At $q=q_\text{crit}$ the 
branches merge into a single one, which remains also at $q>q_\text{crit}$.}
\label{TinvEposk}}


\section{Supersymmetric waves on strings}
\label{waves}

\subsection{Construction of the solution}

We now would like to construct supersymmetric generalizations of the magnetic
string solutions of \cite{Chamseddine:1999xk,Klemm:2000nj}, that carry momentum
along the string. To this end, we recall that
supersymmetric solutions of minimal gauged supergravity in five
dimensions are divided into timelike and null classes, according to
the nature of the Killing vector constructed as a bilinear from the
Killing spinor. The general null solution has been obtained in
\cite{Gauntlett:2003fk} and reads\footnote{In this section, the
  five-dimensional geometries are described by the coordinates
  $\{u,v,x^1,x^2,x^3\}$, where $x^1=z$, $x^2=x$ and $x^3=y$. Latin
  letters $i,j,\dots$ are indices on the three-dimensional flat space
  parameterized by $\{x^1,x^2,x^3\}$. Early greek letters
  $\alpha,\beta,\dots$ are indices of the two-dimensional space
  $\{x^2,x^3\}$, again with flat metric. The antisymmetric tensor
  $\varepsilon_{\alpha\beta}$ on this space is defined such that
  $\varepsilon_{23}=1$, and
  $\Delta^{(2)}=\partial_\alpha\partial_\alpha$ is the flat Laplacian
  in two dimensions.} 
\begin{equation}
\begin{aligned}\label{gsol}
\de s^2&=-H^{-1}(\mathcal{F}\de u^2+2\de u\de v)
+H^2[(\de x^1+a_1\de u)^2+\e^{3\phi}(\de x^{\alpha}+
\e^{-3\phi}a_{\alpha} \de u)^2]\,,\\
A&=A_u \de u
+\frac{\sqrt3}{4g}\varepsilon_{\alpha\beta} \phi_{,\alpha} \de x^{\beta}\,.
\end{aligned}
\end{equation}
The function $\phi(u,x^i)$ is determined by the equation
\begin{equation}\label{gphi}
\e^{2\phi}\partial^2_z \e^{\phi}+\Delta^{(2)}\phi=0\,.
\end{equation}
Given a solution of \eqref{gphi}, $H(u,x^i)$ is obtained from
\begin{equation}
H=-\frac{1}{2g}\phi_{,z}\,,
\end{equation}
and $A_u(u,x^i)$ is found by solving the Maxwell equation
\begin{equation}\label{gAu}
\partial_z[H^2 \e^{2\phi}\partial_z(\e^\phi A_u)]
+\partial_\alpha (H^2 A_{u,\alpha})=
\frac{\sqrt3}{2g} H 
\varepsilon_{\alpha \beta} \phi_{,\alpha u} H_{,\beta}\,.
\end{equation}
Then, the functions $a_i(u,x^j)$ are determined by the system
\begin{equation}
\begin{aligned}\label{ga}
\frac 1{2\sqrt3}\varepsilon_{\alpha\beta}\partial_{\alpha}(H^3 a_\beta)
&=-H^2 \e^{2\phi}
\partial_z(\e^\phi A_u)\,,\\
\frac 1{2\sqrt3}[\partial_\alpha(H^3 a_1)-\partial_z(H^3a_\alpha)]
&= H^2 \varepsilon_{\alpha\beta} A_{u,\beta} 
- \frac{\sqrt3}{4g} H^2 \phi_{,\alpha u}\,,
\end{aligned}
\end{equation}
whose integrability condition is \eqref{gAu}. Finally, the function 
$\mathcal{F}(u,x^i)$ follows from the $uu$-component of the Einstein equations,
\begin{equation}
R_{uu}=2F_{u\sigma}F_u^{\phantom{u}\sigma}-\frac{1}{3}g_{uu}(F^2+12g^2)\,.
\end{equation}
To find the subset of supersymmetric null solutions which describe
waves on strings, we suppose $\phi$ to be separable,
\begin{equation}
\phi(u,x,y,z)=\phi_1(u,z)+\phi_2(u,x,y)\,.
\end{equation}
Substituting this expression of $\phi$ in the equation \eqref{gphi}, we
find that $\phi_1$ and $\phi_2$ have to satisfy the equations
\begin{align}
 \label{phi1}\partial^2_z \e^{\phi_1}&=\frac{k}{24g}\e^{-2\phi_1}\,,\\
\label{phi2}\Delta^{(2)}\phi_{2}&=-\frac{k}{24g}\e^{3\phi_2}\,,
\end{align}
where $k(u)$ is an arbitrary function. \eqref{phi1} implies
\begin{equation}
 \e^{3\phi_1} (\phi_{1,z})^2=\mu \e^{\phi_1}-\frac{k}{12g}\,,
\end{equation}
where $\mu(u)$ denotes again an arbitrary function. Equation \eqref{phi2}
is the Liouville equation. As a particular solution we choose
\begin{equation}
 \e^{3\phi_2}=\frac{64g}{\Upsilon^2}\,,
\end{equation}
where $\Upsilon(u,x,y)=1+k(x^2+y^2)$. To proceed we suppose that $A_u$
is a function of $u$ and $z$ only. Then \eqref{gAu} implies that
\begin{equation}
 A_u= \e^{-\phi_1}\left[\alpha\int \de z\ (\e^{\phi_1} \phi_{1,z})^{-2}+\beta \right]\,,
\end{equation}
with $\alpha(u)$ and $\beta(u)$ arbitrary functions. The system
\eqref{ga} is solved by 
\begin{equation}
Ha_1=\frac{k'}{gk}\frac{\Upsilon-1}{\Upsilon}+\Gamma\,, \qquad
H^3 a_2=\frac{16 \sqrt3  \alpha y}{g\Upsilon}\,, \qquad
H^3 a_3=-\frac{16 \sqrt3  \alpha x}{g\Upsilon}\,,
\end{equation}
where $k'=\partial_u k$ and $\Gamma(u,z)$ is an arbitrary function.
We introduce the new coordinate $\rho$ defined by
\begin{equation}
 \rho=\frac{1}{2\phi_{1,z}(g\e^{\phi_1})^{3/2}}\,,
\end{equation}
and choose
\begin{equation}
 \Gamma=-H\partial_u z\,,
\end{equation}
where $z$ has to be considered as a function of $u$ and $\rho$, and
the derivative has to be taken considering $\rho$ as fixed. 
Then, if we rescale $\de u\rightarrow\mu^{3/2}\de u$,
$\mathcal{F}\rightarrow\mu^{-3/2}\mathcal{F}$ and
$\beta\rightarrow\mu^{-5/2}\beta$ to eliminate $\mu$, in the coordinate
system $\{u,v,\rho,x,y\}$, the general solution \eqref{gsol} takes the form
\begin{equation}
\begin{aligned}\label{gwave}
\de s^2&=\frac{ h^{\frac{3}{2}}}{g^2\rho^2}
\left(\frac{1}{2}\mathcal{F}\de u^2+\de u\de v\right)
+\left(\frac{1}{g\rho h}\de\rho
+\frac{k'}{gk}\frac{\Upsilon-1}{\Upsilon}\de u\right)^2 \\
&\phantom{{}={}}+\frac{4}{g^4\rho^2\Upsilon^2}\left[\left 
(\de x-2\sqrt3g\alpha y\Upsilon h^{\frac{3}{2}} \de u \right)^2 
+ \left(\de y+2\sqrt3g\alpha x\Upsilon h^{\frac{3}{2}} \de u \right)^2\right]
\,, \\
A_u&=\alpha\left\{-2h^{\frac{1}{2}}+3h^{-\frac{1}{2}} +
\frac{g^2 k \rho^2}{h}\ln\left[\frac{1}{g\rho}
\left(1+h^{\frac{1}{2}}\right)\right]\right\}+\frac{4\beta g^3\rho^2}{h}\,,\\
A_x&=\frac{ky}{g\sqrt3\Upsilon}\,,\ \ \ \ A_y=-\frac{kx}{g\sqrt3 \Upsilon}\,,
\end{aligned}
\end{equation}
where 
\begin{equation}
 h=1+\frac{g^2 k \rho^2}{3}\,.
\end{equation}
In the case in which $\alpha=0$ and $k$ is constant, the solution
\eqref{gwave} becomes 
\begin{equation}
\begin{aligned}
\de s^2&=\frac 1{g^2\rho^2 h^2}
\left[\de\rho^2 + h^{7/2}\left(\frac{\mathcal{F}}{2}\de u^2 + \de u\de v\right)
+\frac{4h^2}{g^2}\frac{\de x^2+\de y^2}{\Upsilon^2}\right]\,,\\
A_u&=\frac{4\beta g^3\rho^2}{h}\,,\qquad
A_x=\frac{ky}{g\sqrt3\Upsilon}\,,\qquad 
A_y=-\frac{kx}{g\sqrt3\Upsilon}\,.
\end{aligned}
\end{equation}
This solution describes a string with non-vanishing electric charge density.
The $uu$-component of the Einstein equations gives
\begin{equation}
\Upsilon^2 \Delta^{(2)} \mathcal{F}
+\frac{8 h^2 -20 h}{g^2\rho}\mathcal{F}_{,\rho} 
+\frac{4 h^2}{g^2} \mathcal{F}_{,\rho\rho}
+2048\beta^2 g^6\frac{\rho^4}{h^{7/2}}=0\,.
\end{equation}
A particular solution of this equation is
\begin{equation}
\mathcal{F}
=\frac{3456 \beta^2 g^2}{k^3}
\left(\frac{1}{7 h^{7/2}}-\frac{1}{5 h^{5/2}}\right)\,.
\end{equation}

In the case in which $\alpha=\beta=0$ the solution \eqref{gwave} becomes
\begin{equation}
\begin{aligned}\label{wave}
\de s^2&=\frac{ h^{\frac{3}{2}}}{g^2 \rho^2}
\left(\frac{1}{2}\mathcal{F}\de u^2+\de u\de v\right)
+\left(\frac{1}{g \rho h}\de\rho
+\frac{k'}{g k}\frac{\Upsilon-1}{\Upsilon}\de u\right)^2+\frac{4}{g^4 \rho^2}\frac{\de x^2+\de y^2}{\Upsilon^2}\,,\\
A_u&=0\,,\qquad
A_x=\frac{ky}{g\sqrt3\Upsilon}\,,\qquad 
A_y=-\frac{kx}{g\sqrt3 \Upsilon}\,.
\end{aligned}
\end{equation}
This solution describes a wave (with profile given by $\cal F$) that propagates on a
string in an asymptotically AdS$_5$ space time. 
In this case the $uu$-component of the Einstein equations is
\begin{equation}
\begin{aligned}\label{szeq}
&\Upsilon^2 \Delta^{(2)}\mathcal{F}
+\frac{8 h^2 -20 h}{g^2\rho}\mathcal{F}_{,\rho} +
\frac{4 h^2}{g^2} \mathcal{F}_{,\rho\rho} \\
&+\frac{4\rho^2{k'}^2}{h^{7/2}k}\left(-\frac{14}{3}+\frac{46 h}{3}-16h^2 -
\frac{32 h}{3\Upsilon}+\frac{32 h^2}{\Upsilon}-\frac{16 h^2}{\Upsilon^2}\right)
-\frac{16 \rho^2 k''}{3h^{5/2}}\left(1-2h+\frac{2h}{\Upsilon}\right)=0\,.
\end{aligned}
\end{equation}
A particular solution of this equation is
\begin{equation}
 \mathcal{F}=\mathcal{H}+\mathcal{P}\,,
\end{equation}
where $\mathcal{H}$ is a solution of the homogeneous part of the equation 
\eqref{szeq},
\begin{equation}
\mathcal{H}(u,\rho,x,y)=\mathcal{H}_1(u,\rho)\mathcal{H}_2(u,x,y)\,,
\end{equation} 
where $\mathcal{H}_1$ has to satisfy Heun's equation\footnote{We have 
introduced the new radial coordinate $r=-\frac{1}{3}g^2 k\rho^2$ and
defined $\dot{\mathcal{H}_1}=\partial_r \mathcal{H}_1$.}
\begin{equation}
\ddot{\mathcal{H}}_1+\left(\frac{5/2}{r-1}-\frac{1}{r}\right)
\dot{\mathcal{H}}_1 - \frac{3c/(16k)}{r(r-1)^2}\mathcal{H}_1 = 0\,,
\end{equation}
and $\mathcal{H}_2$ has to obey the Laplace equation
\begin{equation}
\widehat{\triangle}^{(2)}\mathcal{H}_2=c\mathcal{H}_2\,,
\end{equation}
where $\widehat{\triangle}^{(2)}$ is the Laplacian  related to the metric
$\de s^2=(\de x^2+\de y^2)/\Upsilon^2$ and $c$ is a function of  $u$ only.

$\mathcal{P}$ is a particular solution of the equation \eqref{szeq}, 
\begin{equation}
 \mathcal{P}
=\mathcal{A}+\frac{\mathcal{B}}{\Upsilon}+\frac{\mathcal{C}}{\Upsilon^2}\,,
\end{equation}
where
\begin{equation}
\begin{aligned}
\mathcal{A}(u,\rho)
&=\frac{{k'}^2}{g^2 h^{7/2}k^3}\left(\frac 32 - \frac{99h}{10}+
18h^2\right)+\frac{ k''}{g^2 h^{5/2}k^2}\left(\frac{6}{5}-12h^2\right)\,, \\
\mathcal{B}(u,\rho)
&=\frac{{k'}^2}{g^2 h^{5/2}k^3}(6-28h+12h^2)+\frac{4 k''}{g^2 h^{3/2}
k^2}\,,\\
\mathcal{C}(u,\rho)&=\frac{{k'}^2}{g^2 h^{3/2}k^3}(8-6h)\,.
\end{aligned}
\end{equation}

The sections of the geometry \eqref{wave} with constant $u$, $v$ and $\rho$ are
two-dimensional spaces that have curvature  proportional to $k$. It is
amusing to note that, by choosing e.g.\ $k(u)=\tanh u$, these sections
can change continuously from a hyperbolic space $\mathbb{H}^2$ to a sphere
S$^2$ as the coordinate $u$ varies from $-\infty$ to $+\infty$.

In the case $k=-1$ and $\mathcal{F}=0$ the spacetime is described by the metric
\begin{equation}
\de s^2=\frac 1{g^2\rho^2 h^2}\left(\de\rho^2 + h^{7/2}\de u \de v+\frac{4h^2}{g^2}
\frac{\de x^2+\de y^2}{\Upsilon^2}\right)\,.
\end{equation}
This geometry describes a magnetic black string that asymptotes to
AdS$_5$ \cite{Klemm:2000nj}. This solution preserves one quarter of 
supersymmetry and approaches the half-supersymmetric product space
AdS$_3\times\mathbb{H}^2$ near the event horizon at $\rho=\sqrt 3/g$.

\subsection{Siklos-Virasoro invariance}

An interesting property of the full family of solutions \eqref{wave}
is that they enjoy a large reparametrization invariance, called
Siklos-Virasoro invariance. Indeed if we perform the diffeomorphism
\begin{equation}
\overline{u}=\chi(u)\,,\ \ \ \ 
\overline{v}=v-\lambda(u,\rho,x^\alpha)\,,\ \ \ \
\overline{\rho}=\sqrt{\chi'}\rho\,,\ \ \ \ 
\overline{x}^\alpha=\sqrt{\chi'}x^\alpha\,,
\end{equation}
defined by the arbitrary function $\chi(u)$ and
\begin{equation}
\lambda(u,\rho,x^\alpha)=\frac{\chi''}{2\chi'}\left[-\frac 6{5g^2kh^{5/2}} +
\frac 4{g^2k h^{3/2}}\frac{\Upsilon-1}{\Upsilon}\right]+\sigma(u)\,,
\end{equation}
where $\sigma(u)$ is an arbitrary function, the metric and the field
equations remain invariant in form if the functions $k$, $h$,
$\Upsilon$ and $\mathcal{F}$ transform according to
\begin{equation}
\overline{k}=\frac{k}{\chi'}\,,\ \ \ \ 
\overline{h}=1+\frac{g^2\overline{\rho}^2
\overline{k}}{3}\,,\ \ \ \ 
\overline{\Upsilon}=1+\overline{k}(\overline{x}^2 +
\overline{y}^2)\,,
\end{equation}
\begin{equation}
\begin{aligned}
\overline{\mathcal{F}}&=\frac{1}{\chi'}\left\{\mathcal{F}+2{\sigma}' 
+\left[\frac{6k'}{5 g^2 k^2 h^{5/2}}
+\frac{\rho^2k'}{k h^{7/2}}\right]\frac{\chi''}{\chi'}
+\left[\frac{6}{5 g^2 k h^{5/2}}-\frac{\rho^2}{2h^{7/2}}\right]
\left(\frac{\chi''}{\chi'}\right)^2\right.\\
&\phantom{{}={}\frac{1}{\chi'}\left\{\right.}
-\frac{6}{5 g^2 k h^{5/2}}\frac{\chi'''}{\chi'}+\left(\frac{\Upsilon-1}
{\Upsilon}\right)^2\left[\frac{4\rho^2k'}{k h^{3/2}}\frac{\chi''}{\chi'}
-\frac{2\rho^2}{h^{3/2}}\left(\frac{\chi''}{\chi'}\right)^2\right]\\
&\phantom{{}={}\frac{1}{\chi'}\left\{\right.}
+\frac{\Upsilon-1}{\Upsilon}\Bigg[\left(-\frac{4\rho^2
k'}{k h^{5/2}}-\frac{4  k'}{g^2 k^2 h^{3/2}}+\frac{4  k'}{
g^2 k^2 h^{3/2}\Upsilon}\right)\frac{\chi''}{\chi'}\\
&\phantom{{}={}\frac{1}{\chi'}\left\{
+\frac{\Upsilon-1}{\Upsilon}\Bigg[\right.}
+\left.\left.\left(-\frac{4}{g^2 k h^{3/2}}+\frac{2\rho^2}{h^{5/2}}
-\frac{2}{g^2 k h^{3/2}\Upsilon}\right)\left(\frac{\chi''}{\chi'}\right)^2
+\frac{4}{g^2 k h^{3/2}}\frac{\chi'''}{\chi'}\right]\right\}\,.
\end{aligned}
\end{equation}
In particular, 
the equation \eqref{szeq} remains invariant in form under this
transformation. In the case in which $k\rightarrow0$ this invariance
was first obtained by Siklos \cite{Sik}, and if we choose
\begin{equation}
\begin{aligned}
&\sigma(u)=\frac{3}{5g^2k}\frac{\chi''}{\chi'}+\tau(u)\,,\\
&\tau(u)\stackrel{k\rightarrow0}{\longmapsto} 0\,,
\end{aligned}
\end{equation}
then $\mathcal{F}$ transforms in a simple way,
\begin{equation}
\overline{\mathcal{F}}\stackrel{k\rightarrow0}{\longmapsto}\frac{1}{\chi'}
\left[\mathcal{F}+\{\chi,u\}\left(\rho^2+\frac{4}{g^2}(x^2+y^2)\right)\right]
\,,
\end{equation}
where
\begin{equation}
\{\chi(u);u\}
=\frac{\chi'''(u)}{\chi'(u)}-\frac{3}{2}\left(\frac{\chi''(u)}{\chi'(u)}
\right)^2\,,
\end{equation}
denotes the Schwarzian derivative.

It would be interesting to see if this Virasoro symmetry is related to the
microstate counting of black strings in AdS (see \cite{Banados:1999tw} for related work).

\section{Final remarks}

\label{finalrem}

While much effort has been dedicated to the study of the phases of black holes in Minkowski spacetime times a circle, the study of the black hole phases in locally asymptotically AdS$_5$ spacetimes with a non-trivial asymptotic circle is still in its embryonic stages. At the moment, only the uniform black string phase has been investigated, and we generalized this phase to the inclusion of the magnetic charge.
Many interesting questions are still open. First of all, it would be interesting to study the stability of these black strings and to see whether a Gregory-Laflamme instability is present. It is also relevant to study whether more general phases exist, as for example non-uniform strings and localized black holes as in the Kaluza-Klein case. In the presence of a negative cosmological constant we have the additional tool of the dual field theory. By interpreting the asymptotic $\text{S}^1$ as a Scherk-Schwarz circle, the study of lumps of deconfined plasma, using the effective fluid dynamics description, can give us some new insight on different possible coexisting phases. This is currently under investigation.

Also, as we saw, the extremal string solutions with $\mu=0$ interpolate between AdS$_5$ at infinity and AdS$_3$ $\times$ $\cal S$ near the horizon. The field theory dual of this supergravity solution has the interpretation of an RG flow across dimensions:
The ${\cal N}=4$ SYM theory on $\bR$ $\times$ S$^1$ $\times$ $\cal S$ flows to a
two-dimensional conformal field theory in the infrared. We can compute the central charge
of this CFT as follows. According to \cite{Brown:1986nw}, one has
\begin{displaymath}
c = \frac{3R_{\mathrm{AdS}_3}}{2G_3}\,,
\end{displaymath}
where $G_3$ denotes the effective three-dimensional Newton constant related to $G_5$ by
\begin{displaymath}
\frac 1{G_3} = \frac{r_\text{h}^2 V_k}{G_5} = \frac{2 g^3 r_\text{h}^2 V_k N^2}{\pi}\,,
\end{displaymath}
and we used the AdS/CFT dictionary in the last step. The curvature radius
$R_{\mathrm{AdS}_3}$ can be read off from the near-horizon solution \eqref{Fmu=0},
with the result $R_{\mathrm{AdS}_3}=2/\gamma\sqrt{\alpha_2}$. This yields the
central charge
\eq
c = \frac{6 g^3 r_\text{h}^2 V_k N^2}{\pi\gamma\sqrt{\alpha_2}}\,. \label{c}
\feq
In the case where $\cal S$ is a compact Riemann surface of genus $h$ ($k=-1$,
$V_{-1} = 4\pi(h-1)$), and quantized magnetic charge, $q^2=1/12g^2$, \eqref{c}
reduces correctly to the result of \cite{Maldacena:2000mw}, namely $c = 8N^2(h-1)/3$
(cf.~equation~(75) of \cite{Maldacena:2000mw} for $a=1/3$, which is the case of minimal
gauged supergravity).

It would be very interesting to see whether the thermodynamic entropy of near-extremal
black strings in AdS$_5$ can be reproduced by using the central charge \eqref{c}
in Cardy's formula \cite{Banados:1999tw}.

\acknowledgments

We would like to thank O.~Dias for useful discussions.
This work was partially supported by INFN, PRIN prot.~2005024045-002 and by the
European Commission program MRTN-CT-2004-005104.

\normalsize

\appendix

\section{Fefferman-Graham expansion}
\label{appendix}

We give here the first few terms of the Fefferman-Graham expansion of the black strings. For convenience, we define $u=gr$ and $\tilde q=gq$.
Note that $e^{2T}$ and $e^{2V}$ have the same expansion, up to the exchange of $c_z$ and $c_t$, which simply corresponds to the double analytic continuation swapping black strings with bubbles of nothing. Also the $\ln u$ terms always come with a $\xi$ factor.

\begin{equation}
\begin{aligned}
y &= u^2 + \frac23k + \frac{\xi \ln u}{u^2} +\frac{c_z + c_t
    +\frac13k^2\tilde q^2}{u^2} 
    -\frac{k\xi\ln u}{3u^4}-\lp c_z+c_t-\frac{k^2}{24}(1+12\tilde q^2)\rp
    \frac k{3u^4}\\
    &\phantom{{}={}}-\frac{\xi^2}6\frac{(\ln u)^2}{u^6}
    -\lp c_z+c_t-\frac{11}{24}k^2\rp\frac{\xi\ln u}{3u^6}
        \\
    &\phantom{{}={}}\qquad\qquad
    +\lp\frac{11}{12}k^2(c_z+c_t)-4c_zc_t-\frac{23}{576}k^4
    -\frac13k^4\tilde q^2-\frac14k^4\tilde q^4\rp\frac1{6u^6}\\
    &\phantom{{}={}}+\frac{2k\xi^2\lp \ln u\rp^2}{9u^8}
    +\lp c_z+c_t-\frac{13}{72}k^2-\frac1{12}k^2\tilde q^2\rp\frac{4k\xi\ln u}{9u^8}
        \\
    &\phantom{{}={}}\qquad    
    +\lp c_z^2+c_t^2+6c_zc_t-\frac{k^2}3\lp\frac{13}{6}
    +\tilde q^2\rp(c_z+c_t)
    +\frac{77}{2592}k^4+\frac{91}{432}k^4\tilde q^2+\frac7{36}k^4\tilde q^4
    \rp\frac k{9u^8}
    \\
        &\phantom{{}={}}+\frac{\xi^3(\ln u)^3}{12u^{10}}
        + O\left(\frac{(\ln u)^2}{u^{10}}\right)\,, 
\end{aligned}
\end{equation}
\begin{equation}
\begin{aligned}
e^{2T} &= u^2 + \frac k2  
        + \frac{\xi \ln u}{2u^2} +\frac{c_z}{u^2} 
    -\frac{7k\xi\ln u}{36u^4}
    -\lp c_z+\frac34c_t-\frac{11}{288}k^2-\frac7{24}k^2\tilde q^2\rp
    \frac {2k}{9u^4}\\
    &\phantom{{}={}}-\xi^2\frac{(\ln u)^2}{8u^6}
    -\lp c_z+c_t-\frac{29}{72}k^2+\frac16k^2\tilde q^2\rp\frac{\xi\ln u}{4u^6}
    +\lp\frac{59}{576}k^2c_z+\frac{19}{192}k^2c_t-\frac12c_zc_t
\right.    
       \\
    &\phantom{{}={}}\qquad\qquad\qquad\qquad\qquad
\left.
    -\frac{199}{41472}k^4-\frac1{48}k^2c_z\tilde q^2-\frac1{16}k^2c_t\tilde q^2
    -\frac5{216}k^4\tilde q^2-\frac1{32}k^4\tilde q^4\rp\frac1{u^6}\\
    &\phantom{{}={}}+\frac{127}{720}k\frac{\xi^2\lp \ln u\rp^2}{u^8}
    +\lp \frac{25}{72}c_z+\frac{43}{120}c_t-\frac{2591}{43200}k^2+\frac{17}{1200}k^2\tilde q^2\rp\frac{k\xi\ln u}{u^8}
        \\
    &\phantom{{}={}}\qquad
    +\lp \frac{4}{45}c_z^2+\frac1{10}c_t^2+\frac{31}{60}c_zc_t
    -\frac{1027}{17280}k^2c_z-\frac{581}{9600}k^2c_t
    \right.
           \\
    &\phantom{{}={}}\qquad\qquad\qquad
\left.    
    +\frac{1}{160}k^2c_z\tilde q^2+\frac{53}{2400}k^2c_t\tilde q^2
    +\frac{250849}{93312000}k^4+\frac{9989}{972000}k^4\tilde q^2\rp\frac k{u^8}
    \\
        &\phantom{{}={}}+\frac{11\xi^3(\ln u)^3}{144u^{10}}
        + O\left(\frac{(\ln u)^2}{u^{10}}\right)\,, 
\end{aligned}
\end{equation}
\begin{equation}
\begin{aligned}
e^{2V} &= u^2 + \frac k2  
        + \frac{\xi \ln u}{2u^2} +\frac{c_t}{u^2} 
    -\frac{7k\xi\ln u}{36u^4}
    -\lp c_t+\frac34c_z-\frac{11}{288}k^2-\frac7{24}k^2\tilde q^2\rp
    \frac {2k}{9u^4}\\
    &\phantom{{}={}}-\xi^2\frac{(\ln u)^2}{8u^6}
    -\lp c_z+c_t-\frac{29}{72}k^2+\frac16k^2\tilde q^2\rp\frac{\xi\ln u}{4u^6}
    +\lp\frac{59}{576}k^2c_t+\frac{19}{192}k^2c_z-\frac12c_zc_t
\right.    
       \\
    &\phantom{{}={}}\qquad\qquad\qquad\qquad\qquad
\left.
    -\frac{199}{41472}k^4-\frac1{48}k^2c_t\tilde q^2-\frac1{16}k^2c_z\tilde q^2
    -\frac5{216}k^4\tilde q^2-\frac1{32}k^4\tilde q^4\rp\frac1{u^6}\\
    &\phantom{{}={}}+\frac{127}{720}k\frac{\xi^2\lp \ln u\rp^2}{u^8}
    +\lp \frac{25}{72}c_t+\frac{43}{120}c_z-\frac{2591}{43200}k^2+\frac{17}{1200}k^2\tilde q^2\rp\frac{k\xi\ln u}{u^8}
        \\
    &\phantom{{}={}}\qquad
    +\lp \frac{4}{45}c_t^2+\frac1{10}c_z^2+\frac{31}{60}c_zc_t
    -\frac{1027}{17280}k^2c_t-\frac{581}{9600}k^2c_z
    \right.
           \\
    &\phantom{{}={}}\qquad\qquad\qquad
\left.    
    +\frac{1}{160}k^2c_t\tilde q^2+\frac{53}{2400}k^2c_z\tilde q^2
    +\frac{250849}{93312000}k^4+\frac{9989}{972000}k^4\tilde q^2\rp\frac k{u^8}
    \\
        &\phantom{{}={}}+\frac{11\xi^3(\ln u)^3}{144u^{10}}
        + O\left(\frac{(\ln u)^2}{u^{10}}\right)\,.
\end{aligned}
\end{equation}

\end{document}